\documentclass[12pt,preprint]{aastex}

\shorttitle{Molecular Evolution in Collapsing Prestellar Cores III}
\shortauthors{Aikawa et al.}

\begin{document}


\title{Molecular Evolution in Collapsing Prestellar Cores III:\\
Contraction of A Bonnor-Ebert Sphere}


\author{Yuri Aikawa}
\affil{Department of Earth and Planetary Sciences, Kobe University,
Kobe 657-8501, Japan}

\author{Eric Herbst}
\affil{Departments of Physics, Chemistry, and Astronomy, The Ohio
State University,
Columbus, OH 43210, USA}

\author{Helen Roberts\altaffilmark{1}}
\affil{Department of Physics, The Ohio State University,
Columbus, OH 43210, USA}

\and

\author{Paola Caselli}
\affil{Osservatorio Astrofisco di Arcetri, Largo Enrico Fermi 5, I-50125
Firenze, Italy}

\altaffiltext{1}{Present address: Department of Physics, UMIST, PO Box 88,
Manchester M60 1QD, UK}

\begin{abstract}
The gravitational collapse of a spherical cloud core is investigated
by numerical calculations.  The initial conditions of the core lie
close to the critical Bonnor-Ebert sphere with a central density of
$\sim 10^4$ cm$^{-3}$ in one model ($\alpha=1.1$),
while gravity overwhelms pressure in the other ($\alpha=4.0$), where
$\alpha$ is the internal gravity-to-pressure ratio. The
$\alpha=1.1$ model shows reasonable agreement with the observed
velocity field in prestellar cores.  Molecular distributions in cores
are calculated by solving a chemical reaction network that includes
both gas-phase and grain-surface reactions.  When the central density
of the core reaches $10^5$ cm$^{-3}$, carbon-bearing species are
significantly depleted in the central region of the $\alpha=1.1$
model, while the depletion is only marginal in the other model.  The
two different approaches encompass the observed variations of
molecular distributions in different prestellar cores, suggesting that
molecular distributions can be probes of contraction or accumulation
time scales of cores. The central enhancement of the
NH$_3$/N$_2$H$^+$ ratio, which is observed in some prestellar cores,
can be reproduced under certain conditions by adopting recently
measured branching fractions for N$_2$H$^+$ recombination.
Various molecular species, such as CH$_3$OH and CO$_2$, are produced
by grain-surface reactions. The ice composition depends sensitively on the
assumed temperature.
Multi-deuterated species are included in our most recent gas-grain
chemical network.  The deuterated isotopomers of H$_3^+$ are
useful as probes of the central regions of evolved cores, in which
gas-phase species with heavy elements are strongly depleted.
At 10 K, our model can reproduce the observed abundance ratio of
ND$_3$/NH$_3$, but underestimates the isotopic ratios of deuterated to normal
methanol.

\end{abstract}




\section{Introduction}

Theoretical models show that molecular abundances in molecular clouds
are time-dependent \citep[e.g.][]{ph80}. Taking advantage of this dependence,
the ages of cloud cores have been estimated. For example, cores with a high
abundance of CCS and carbon chain species are considered to be young, while
those with large abundances of NH$_3$ are considered old
\citep{su92}. There have been arguments, however, on the reliability of such
``chemical clocks'', because molecular abundances depend not only on
the elapsed time but also on density, temperature, and the C/O ratio in the
gas phase \citep[e.g.][]{prat1997,th1998}. Moreover, single-dish observations,
by which radio astronomers used to estimate molecular abundances, average cloud
structure and abundances over a beam size.

During the last decade, independent and spatially-resolved estimates
of molecular abundance, temperature, and density have become
available.  Observations of sub-millimeter continua and extinctions of
background stars have revealed density distributions in a variety of starless
dense cores \citep{wt1994,all2001}, showing that the density is almost
constant inside a radius of several thousand AU, and decreases outside
of this radius as $r^{-p}$, where $p$ is $2.0-2.5$ and $r$ is the distance
from the core center.  Spatial distributions of molecular line intensities have
been obtained by interferometric observations and detailed mapping via
single dish telescopes.  Intensity ratios of different transitions
show that the starless cores are almost isothermal with $T\sim 10$ K
\citep{taf1998,taf2002,taf2004}, although a slight deviation of a few
K may exist \citep{ev2001, gal2002}.  Spatial distributions of molecular
abundances within selected cores have been obtained by analyses 
that include
radiative transfer, showing that CO, CS and CCS are depleted at the
core center, while NH$_3$ and N$_2$H$^+$ maintain relatively constant
abundances \citep{cas1999,oha1999,bgn2001,cas2002a, taf2002, taf2004}.
The central depletion indicates that adsorption of
gas-phase species onto grain occurs, a process that had long been
anticipated theoretically \citep{w76, bh1983, ta87}.  Since adsorption lowers
gas-phase molecular abundances as a function of time, it can be used
as another indication of the history of the source \citep{bl1997}.

Considering that dense starless cores should be formed from lower
density gas at $10^3-10^4$ cm$^{-3}$, we can estimate the contraction
time scale via chemical models that include adsorption.
\citet{aik2001} and \citet{aik2003} investigated the distribution of
molecules in contracting cores by adopting a Larson-Penston (L-P) flow
and analogues with slower rates of contraction.  They found the
central depletion to be greater in models with slower rates of
contraction.  The L-P model better reproduces observed molecular
column densities, while the radial distributions of molecules are
better reproduced by a contraction that is three-fold slower.  Since
predicted molecular column densities vary as contraction proceeds,
\citet{aik2003} were able to use their results to sort observed cores
into an evolutionary sequence.  Mono-deuterated species were also
included in their chemical model.  They obtained very high D/H ratios
such as H$_2$D$^+$/H$_3^+ > 1$, which is consistent with the recent
observation of H$_{2}$D$^{+}$ in the core L1544 \citep{cas2003}.  The
L-P flow, however, is an asymptotic solution, which appears in the
central region of a collapsing core.  The collapse velocity at outer
radii, which is of interest in the comparison with observation, is
overestimated in the L-P flow. In their models with slower rates
of contraction, \citet{aik2003} adopted a constant factor $f$ to
decrease the contraction speed.  In reality,  the factor $f$ varies
with time. To avoid this problem, \citet{li2002}
and \citet{sh2003} directly coupled calculations of molecular
evolution with magnetized core collapse.  They showed that the
molecular column densities in L1544 are well reproduced by the
magnetized model.  Although the infall velocities obtained in their
models are generally faster ($\gtrsim 0.2$ km s$^{-1}$) than observed
in L1544 ($\sim 0.1$ km s$^{-1}$), the magnetized model gives a
somewhat smaller, and thus better, infall velocity than the
non-magnetized model.
They also suggested that the molecular line widths will
vary depending on the spatial distribution of the abundances
and radial velocity.

The present paper is an update of \citet{aik2003}. We now
calculate molecular distributions while solving for the contraction of
a spherical core numerically, rather than by adopting an L-P flow, so
that our infall velocities can be compared with observed values.  For
simplicity, we restrict ourselves to the non-magnetized case;
this simplification enables us to separate the dynamics and chemistry
and thus to utilize a large chemical network.
The infall speed is dependent on the ratio between the gravitational and
pressure forces, for which we consider two different cases.  We will
show that observed infall velocities are reproduced, without the help
of a magnetic field, if the initial condition of the core is close to
the critical Bonnor-Ebert sphere \citep{bon1956} with central density
$\sim 10^4$ cm$^{-3}$. On the other hand, magnetic fields are
probably needed in more massive or less pressurized cores to maintain
the infall speed close to observed values. Our newly calculated
molecular distributions are presented in selected stages of contraction
and compared with recent observations of prestellar cores such as L1498,
L1517B, L1521E, and L1544. As shall be shown, our model reproduces some
observed features including an enhancement of the NH$_3$/N$_2$H$^+$
ratio in the central regions, which was not explained in previous
studies.

Our basic chemical model contains a large gas-phase network
\citep[e.g.][]{th1998} that includes mono-deuterated species, and
grain-surface reactions. In addition, we have extended the network
for some calculations to include multi-deuterated species
\citep{rhm2003}. In recent years several multi-deuterated species
have been detected in prestellar and protostellar cores, and their
relative abundances to normal isotopomers, known as D/H ratios,
found to be very high
($10^{-3}-10^{-1}$) \citep{vdt2002, lis2002, par2002, bac2003, par2004}.
Since the D/H ratio increases significantly as density increases and as
gas-phase CO is depleted, it is of interest to see if the abundances
of multi-deuterated species can be reproduced in our contracting core
models.

The remainder of the paper is organized as follows.  A description
of the model is given in \S 2. Numerical results are presented and
compared with observations in \S 3. In \S 4, we summarize the 
chemical and kinematic properties of some prestellar cores, and briefly
discuss the formation of cores and the role of turbulence. A summary is
contained in \S 5.

\section{Model}
\subsection{Contraction of Cores}
We solve the dynamical collapse of a cloud core following \citet{otn1999}.
With the assumption of spherical symmetry, the basic equations are
described in spherical coordinates by
\begin{equation}
\frac{\partial \rho}{\partial t} + \frac{1}{r^2} \frac{\partial}{\partial r}
(r^2 \rho v ) = 0
\end{equation}
\begin{equation}
\frac{\partial}{\partial t}(\rho v) + \frac{1}{r^2}
\frac{\partial (r^2 \rho v^2)}{\partial r} = - \frac{\partial p}{\partial r}
+ \rho g
\end{equation}
and
\begin{equation}
g = -\frac{4\pi G}{r^2} \int_{0}^{r} \rho r^{' 2} dr^{'}
\end{equation}
where
\begin{equation}
p=\rho c_{\rm s}^2.
\end{equation}
    In equations (1)-(4), $t$
is the time, $r$ is the distance from the center of the core, $\rho$
is the mass density, $v$ is the radial velocity, $g$ is the
acceleration due to gravity, $G$ is the gravitational constant, $p$
is the pressure, and $c_{\rm s}$ is the speed of sound.  The gas is
assumed to be isothermal with $T=10$ K.

The equations are solved numerically by the MUSCL-TVD method 
\citep{vl77, vl79, har84}, with the numerical flux calculated by
the Roe method.

The radial distribution of density in an equilibrium core
(i.e., a Bonnor-Ebert sphere) is given by the equations
   \begin{equation}
   \frac{\partial (\rho {c_s}^2)}{\partial r} =  \rho g
   \end{equation}
   and
\begin{equation}
   \frac{\partial (r^2 g)}{\partial r} = - 4 \pi G \rho r^2.
   \end{equation}
The central density of the equilibrium core is set to be $n_{\rm H} =2\times
n({\rm H}_2)+n({\rm H}) = 2\times 10^4$ cm$^{-3}$. Away from the center,
the density is almost constant within $\sim 10000$ AU, and decreases outwards
when $r\gtrsim 10000$ AU.

To ensure contraction, the initial density distribution is determined
by  multiplying the equilibrium density distribution by a constant factor
$\alpha$, equal to the ratio of the gravitational force to the pressure force.
The initial radial velocity is set to zero all through the core.
We consider two cases: $\alpha=1.1$ and 4.0. The former model simulates the
contraction of a nearly-equilibrium core, while the latter case is similar to
the L-P flow, which was adopted in \citet{aik2003}. A fixed boundary
condition ($v=0$) is adopted at the outer core radius of 0.2 pc, which is
slightly larger than the critical radius of an isothermal
equilibrium sphere \citep[see also][]{otn1999}.
The mass of the equilibrium core ($\alpha=1.0$) is then 4.0 $M_{\odot}$,
which is slightly larger than the critical mass (3.9 $M_{\odot}$) of a sphere
with central density $n_{\rm H}=2\times 10^4$ cm$^{-3}$ and kinetic temperature
10 K. Note that by the word ``equilibrium'', we refer to
a density distribution governed by equations (5) and (6); such a distribution,
however, is only stable up to a certain critical mass \citep{bon1956}.

\subsection{Chemistry}
We follow migrating fluid parcels in the contracting cores and solve a
chemical reaction network to obtain the molecular evolution in each
parcel. The chemical reaction network includes gas-phase reactions,
adsorption onto and desorption from dust, and diffusive grain-surface
reactions. The gas-phase network is based on the New Standard Model (NSM)
\citep[e.g.][]{th1998}
but includes mono-deuterated species. We adopt newly measured branching
fractions for the N$_2$H$^+$ recombination in the gas phase.  Previously
it had been assumed that N$_2$H$^+$ dissociatively recombines
to produce N$_2$ + H as the sole products.  Recently, however,
\citet{gep2004} measured the products of the recombination to be NH +
N and N$_2$ + H with branching fractions of 65\% and
35\%, respectively; these values are adopted in our current network.

All neutral species are assumed to be adsorbed when they hit the grain
surface with a sticking probability $S$ of unity.  The adsorbed
species return to the gas phase via thermal desorption and impulsive
heating of the grains by cosmic-rays \citep{ljo1985,hh1993}. We have
adopted the adsorption energies of \citet{hh1993} but replaced them
with experimental data for pure ices when available \citep[see Model B
of][]{aik2001}.
One exception is CO, for which we assumed a binding
energy of 1780 K, which corresponds to adsorption onto polar ice,
instead of the value of pure CO ice (960 K).  With the 
assumption of the lower value, the gas-phase CO abundance was 
calculated to be rather high compared with observed values 
\citep[see the discussion in ][]{aik2003}.

A set of grain-surface reactions was taken from \citet{rh2000}.
The reaction rate coefficients for the diffusive grain-surface reactions
were calculated following \citet{hh1993} as modified by \citet{sch2001}
and \citet{cas2002b}.
The unmodified rate coefficients are  proportional to the sum of the rates
for the two reactants to diffuse over the entire grain \citep{hh1993}.
Activation energies against diffusion are obtained as a constant fraction
(30 \%) of adsorption energies, and the diffusion rate is given by the larger
of the tunneling and thermal hopping rates. If, for reactions 
without chemical activation energy, the calculated
diffusion rates are larger than both the accretion and evaporation rates,
they are replaced by the larger of the latter two. The dominant alteration is
to slow down the reaction rates of atomic hydrogen and deuterium.
The so-called modified rate method used here has been found to be in good
agreement for most species with a more detailed stochastic method
\citep{sh2004}. Note that the diffusion barriers and adsorption
energies used here are different from those used in the gas-grain
models of \citet{rh2000,rh2001}, where higher values were chosen,
based on an extrapolation of the analysis of \citet{ka1999}.

The present paper includes two corrections to the
grain-surface reaction rates adopted in our previous paper \citep{aik2003}.
Firstly, we now estimate the tunneling
rate from activation energies against diffusion $E_{\rm b}$ using a
simple rectangular barrier \citep[e.g.][]{her93}.
In \citet{aik2003} the tunneling rates of some species (H, D, H$_2$, HD, He,
C, N, and O) were estimated using the width of the lowest energy
band $\triangle E$; the rate is $\pi \triangle E / 2 h$, where $h$ is the
Planck constant \citep{ta87}. The former gives slower tunneling rates
than the latter.
Secondly, if a reaction between two minor surface species has a 
chemical activation barrier, we now compare
the total reaction rate coefficient, instead of the diffusion rate, with the 
larger
of the accretion and evaporation rates. For reactions between 
reactive accreting species (e.g. H) and weakly reactive major surface 
species (e.g. CO) with an activation energy barrier,  we utilize a 
modified version of an additional correction proposed by 
\citet{cas2002b} if the surface abundance of the weakly reactive 
species accumulates to the point that the probability of reaction 
exceeds unity in the shorter of the accretion and evaporation 
intervals.  The modification is to replace the standard rate law with 
one containing the accretion rate of the accreting species (or the 
evaporation rate if larger) multiplied by the surface abundance of 
this species and the probability that it reacts with the particular 
weakly reactive species of interest.    Consider, for example, the 
reaction of H atoms with CO and H$_{2}$CO molecules in the limit when 
both have sufficiently large abundances to turn on the correction, 
and suppose that the surface abundance of CO is nine times that of 
formaldehyde.  Since the activation energy barriers are the same, the 
probability that H reacts with CO (0.9) is nine times the probability 
that it reacts with formaldehyde (0.1).  Without the probability 
term, the reaction rates for H with both CO and H$_{2}$CO are the 
same since the abundance of the weakly reactive species is no longer 
in the rate law.   Because it is very difficult to incorporate the 
probability term into a full-scale gas-grain model, we have chosen to 
neglect it.  The result is difficult to gauge but, in any case, the 
use of the modification at this level is certainly preferable to 
ordinary rate equations \citep{sh2004}.  
We have used the 
modification for O + CO and for the reactions of H and D atoms with
CO, C$_2$H$_6$, CH$_4$, H$_2$CO, H$_2$S, and their deuterated 
counterparts. All of these reactions except for those involving CO 
and H$_{2}$CO are hydrogen abstraction processes, in which the 
reacting H atom removes an atom from the other reactant.

Deuterium exchange reactions in the gas phase are based on the work of
\citet{mbh1989}.  In including deuterium fractionation on
dust particle surfaces, we have assumed that the adsorption energy and
barrier against diffusion for surface D atoms are slightly higher than
the values for H atoms \citep{cas2002b}.  In total, our chemical
network consists of 878 species and 11779 gas-phase and surface
reactions.  In \S 3.4.2, we present results based on a new chemical
network extended to include multi-deuterated species \citep{rob2004}.
Since we are
interested in the D/H ratios of specific species, such as H$_2$CO and
NH$_3$, and since a calculation of the fully-multi-deuterated network
is time-consuming, complex species such as long carbon chains have
been excluded from this network, which consists of 518 species and
12616 reactions.

We have utilized the so-called ``low-metal'' values for initial
gas-phase elemental abundances \citep[see Table 1 of][]{aik2001} and
have adopted the ``standard'' value of $\zeta = 1.3 \times 10^{-17}$
s$^{-1}$ for the ionization rate by cosmic rays.  Considering that the
cores we are modeling are embedded in molecular clouds, we assume the
visual extinction $A_{\rm v}$ to be 3 mag at the outer boundary of the
core, so that photodissociation does not significantly affect our
results.  All heavy elements are assumed to be initially in atomic and
ionized form, with carbon ionized and oxygen neutral.  The initial
form of hydrogen is molecular, and the deuterium is assumed to be in
the form of HD, at a ratio of $3.0 \times 10^{-5}$ with respect to
H$_{2}$.

\section{Results}
\subsection{Core Contraction}
\begin{figure}
\plotone{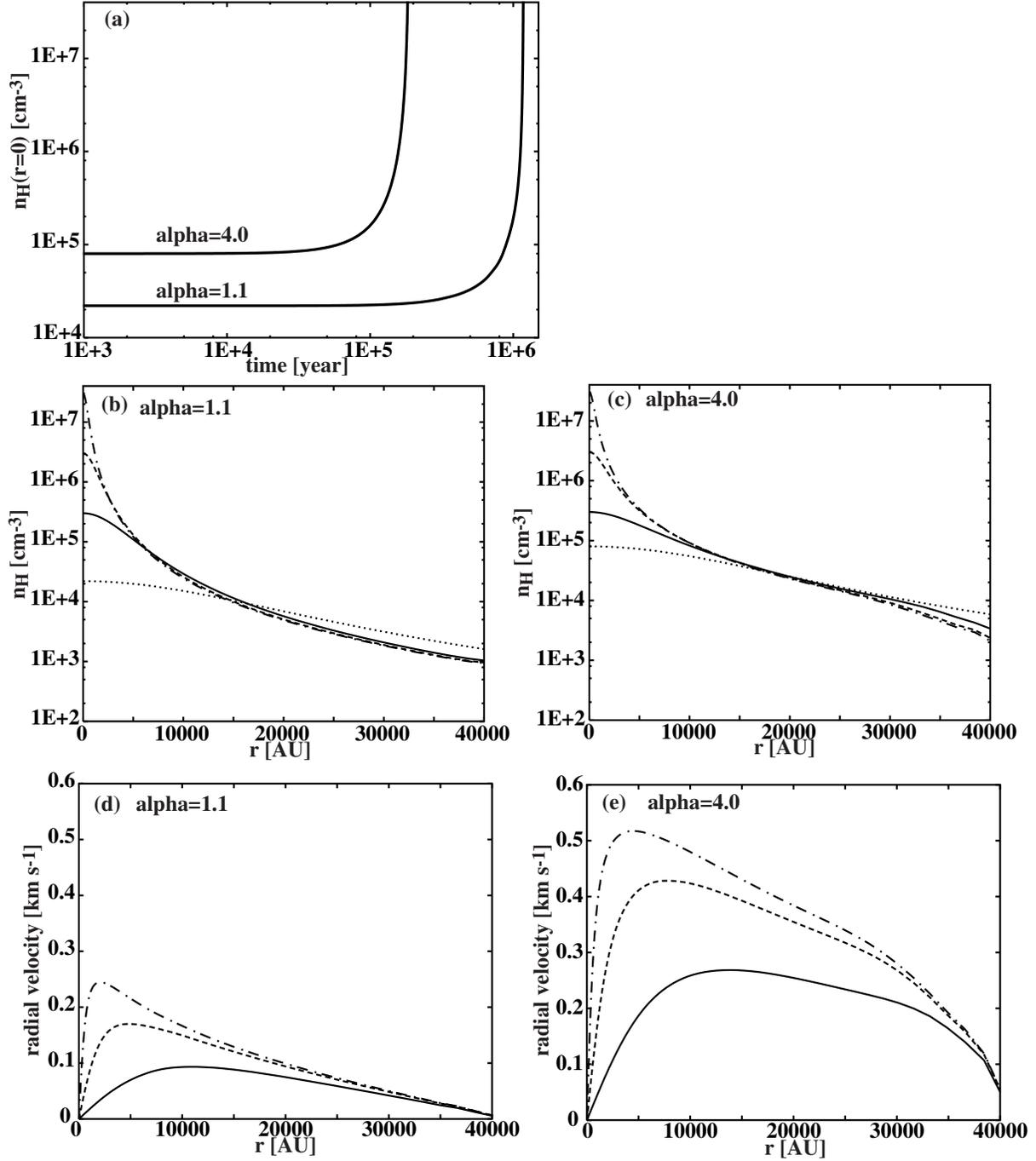}
\caption{The temporal variation of the central number density of the
model cores is shown in (a).
The initial densities of the model cores are larger than the equilibrium
values by a factor of $\alpha$. The density and velocity distributions in the
model with $\alpha=1.1$ are shown in  panels (b) and (d), respectively.
The dotted line represents the initial density distribution, while the solid,
dashed and dot-dashed lines represent the density and velocity distributions
at $t=1.05 \times 10^6$, $1.15 \times 10^6$, and $1.17 \times 10^6$ yr.
The density and velocity distributions in the model with $\alpha=4.0$ are shown
in panels (c) and (e), respectively. The dotted line represents the initial
density distribution, while the solid, dashed and dot-dashed lines represent
the density and velocity distributions at $t=1.29 \times 10^5$, $1.70 \times
10^5$, and $1.82 \times 10^5$ yr.
\label{dist_phys}}
\end{figure}

Figure \ref{dist_phys} (a) shows the temporal variation of the central
density. The central densities of the cores with $\alpha=1.1$ and 4.0 reach
$3\times 10^7$ cm$^{-3}$ at ages of $1.17 \times 10^6$ yr and
$1.82\times 10^5$ yr, respectively. In the former case, the central density
reaches $8\times 10^4$ cm$^{-3}$ (initial density of the $\alpha=4.0$ model)
at $\sim 8.45\times 10^5$ yr and  takes
another $3.27\times 10^5$ yr to reach a density of $3\times 10^7$ cm$^{-3}$.
The core with the larger $\alpha$ contracts faster, because  gravity overwhelms
the thermal pressure.

Distributions of number density and infall velocity at the
initial central density and when the central density is $3\times 10^5,
3\times 10^6$, and $ 3\times 10^7$ cm$^{-3}$ are shown in Figure
\ref{dist_phys} ($b-e$).
In these panels, the density and
velocity are plotted vs $r$ rather than vs log $r$. If the density
distribution had been plotted as a function of log $r$, the density would
appear to be almost constant at $r\lesssim r_{\rm flat}$ and
to decrease as $n_{\rm H}\propto r^{-p}$ at larger radii.  The exponent
$p$ is  $\sim 2.3$ for $\alpha=1.1$ and $\sim 1.9$ for $\alpha=4.0$.
The model with the larger $\alpha$ has the larger $r_{\rm flat}$. When
the central density is $n_{\rm H}= 3\times 10^5$ cm$^{-3}$, for example,
$r_{\rm flat}$ is about $4.0\times 10^3$ AU and $5.4\times 10^3$ cm$^{-3}$
for $\alpha=1.1$ and 4.0, respectively.

The infall velocity is naturally zero at the core center, because there is
no point source of gravity at $r=0$. It is also zero at the outermost
radius (0.2 pc, or $4.1 \times 10^4$ AU) as a fixed boundary value.
The infall velocity hence has a peak value at a certain radius, which
moves inwards as the contraction proceeds.
At the same central density, the model with the larger $\alpha$ has the larger
infall velocities.
It should be noted that if
the contraction starts from a nearly-equilibrium state such as given by
$\alpha=1.1$,  the infall velocity when the central density is $\sim
10^5$ cm$^{-3}$ is in reasonable agreement with the observed values of
$0.05-0.1$ km s$^{-1}$ of some prestellar cores \citep{lee2001, taf2004}.
Although the peak infall velocity (0.093 km s$^{-1}$) with $\alpha=1.1$
and a central density of $3\times 10^5$ cm$^{-3}$ is close to the upper
boundary of the observed values ($\sim 0.1$ km s$^{-1}$), it could be smaller
with a smaller $\alpha$.
For example, the peak infall velocity is 0.073 km s$^{-1}$ at the same central
density if $\alpha$ is 1.05. In addition, the observation may underestimate
the infall velocity by a factor of 1-2, because of morphological effects
\citep{lee2001}.

Since the non-zero infall velocities increase as time progresses,
the contraction can be said to accelerate with time. In the model
with $\alpha=1.1$, the peak value of the infall velocity becomes
comparable to that of the L-P flow (i.e. three times the sound velocity)
at a central density of $10^{17}$ cm$^{-3}$
\citep[Figure 3 of][]{otn1999}. Because the contraction speeds up,
the model is different from the L-P analogues with constant
delaying factor $f$ \citep{aik2003}; the current model corresponds to
the case in which $f$ decreases with time.

The infall velocities in our model with $\alpha=1.1$ tend to be smaller
than those of \citet{li2002} and \citet{sh2003}, in spite of the fact
that our model does not include magnetic support. The difference seems
to be caused by a combination of factors including the mass, size, and
initial density of the model cores.
Since their model cores are initially supported by thermal {\it and} magnetic
pressure, and extend to $10^5$ AU, they are more massive (20 $M_{\odot}$) than
our model with $\alpha=1.1$ (4.4 $M_{\odot}$). The importance of mass
is discussed by \citet{sh2003}, who show that infall
velocities in their 10 $M_{\odot}$ model are smaller than those of their 20
$M_{\odot}$ model. The initial central density of \citet{li2002} is $10^3$
cm$^{-3}$, while it is $\sim 10^4$ cm$^{-3}$ in our models. If we lower
our initial central density to $10^3$ cm$^{-3}$, the peak value of the infall
velocity is 0.16 km s$^{-1}$ when the central density reaches
$2\times 10^5$ cm$^{-3}$.  This velocity is only slightly smaller than
the value (0.2 km s$^{-1}$) of \citet{li2002}.
The boundary condition at the outermost radius also affects the infall
velocity. While we have adopted a fixed boundary condition, \citet{li2002}
adopted a free-pressure boundary condition, which yields larger infall
velocities.

It is interesting that the infall velocities of our core model are similar
to, although slightly larger than, those derived in the MHD model of
\citet{cb2000}, which show good agreement with the observed line width in
L1544 \citep{cas2002c}. Specifically, when their central density reaches
values of  $4\times 10^5$ cm$^{-3}$, $4\times 10^6$ cm$^{-3}$,
and $4\times 10^7$ cm$^{-3}$, the peak infall velocity of their
neutral component is 0.1 km s$^{-1}$ (at $\sim 1.3\times 10^4$ AU),
0.14 km s$^{-1}$ (at $\sim 5.0 \times 10^3$ AU), and 0.17 km s$^{-1}$ (at
$\sim 2.0 \times 10^3$ AU), respectively.

\subsection{Molecular Distributions}
\begin{figure}
\plotone{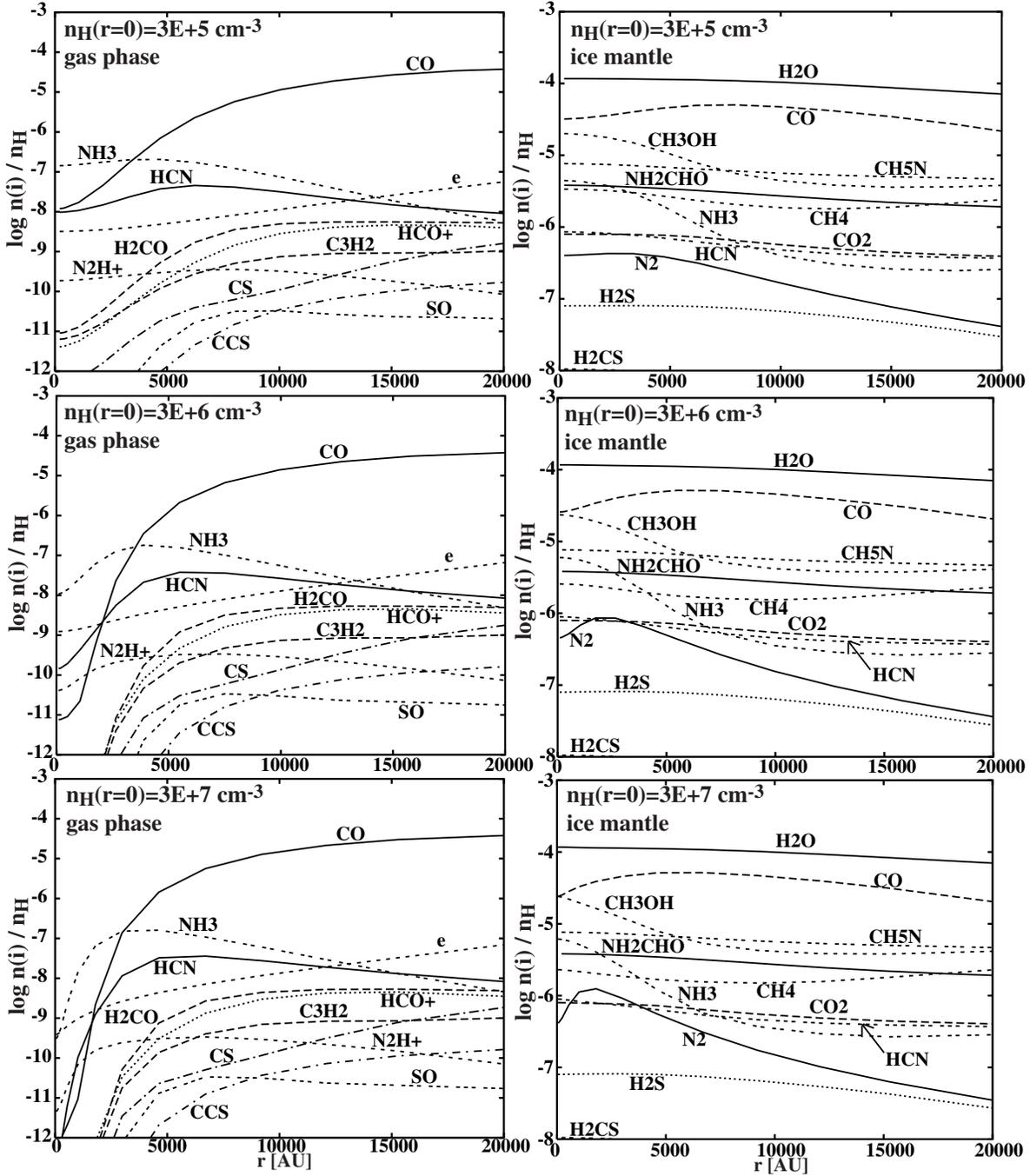}
\caption{Distributions of calculated molecular abundances with $\alpha=1.1$.
The left panels show the gas-phase species, and the right panels show the
ice-mantle species. The central densities are $3\times 10^5$ cm$^{-3}$
(top), $3\times 10^6$ cm$^{-3}$ (middle), and $3\times 10^7$ cm$^{-3}$
(bottom).
\label{dist_abun_alp1}}
\end{figure}

\begin{figure}
\plotone{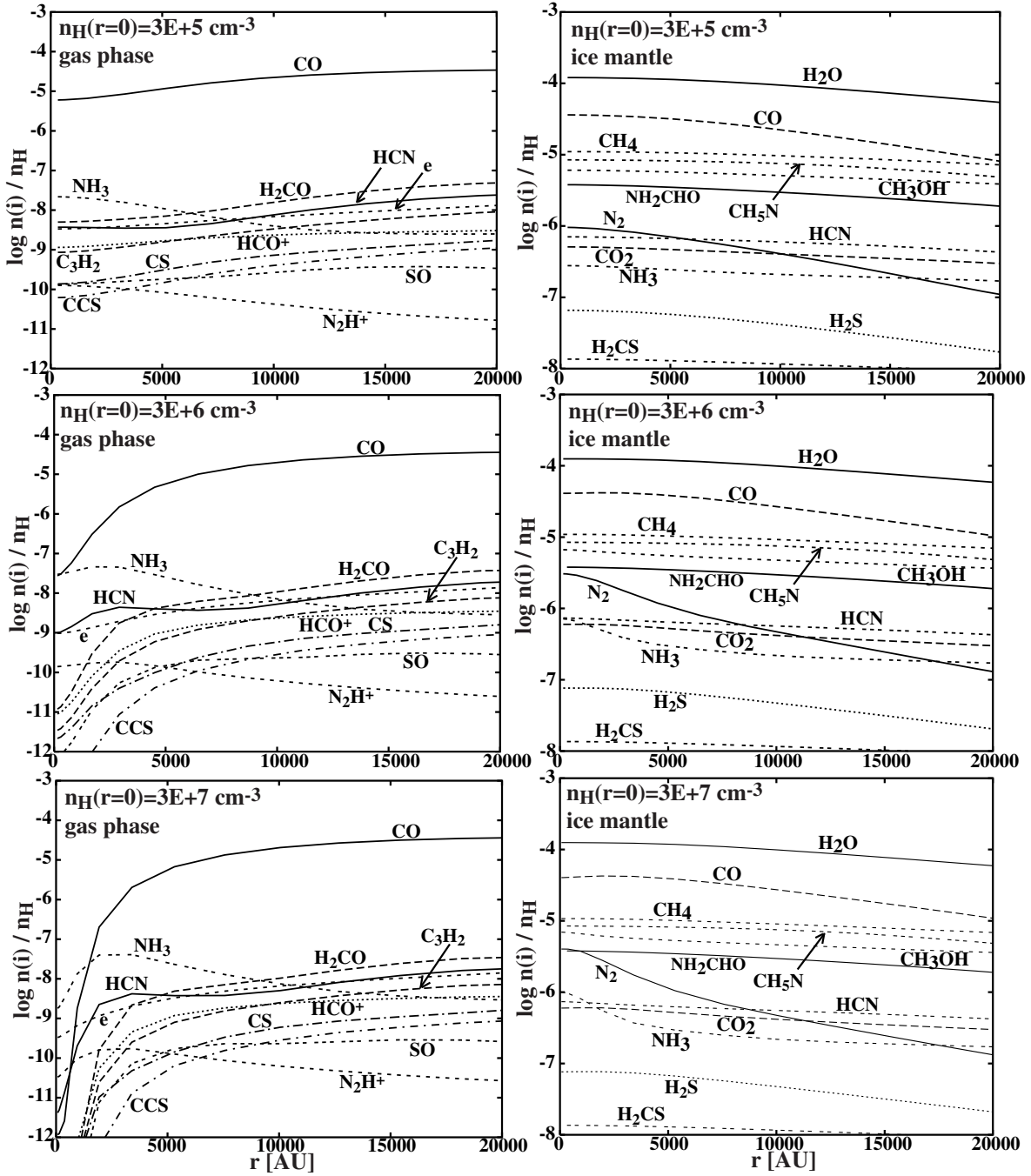}
\caption{Distributions of molecular abundances as in Figure
\ref{dist_abun_alp1}
but for  $\alpha=4.0$.
\label{dist_abun_alp4}}
\end{figure}

Figure \ref{dist_abun_alp1} and Figure \ref{dist_abun_alp4} show
calculated distributions of molecular (fractional) abundances in 
contracting cores
when the central density is $3\times 10^5$ cm$^{-3}$ (top), $3\times
10^6$ cm$^{-3}$ (middle), and $3\times 10^7$ cm$^{-3}$ (bottom) for
the models with $\alpha=1.1$ and 4.0, respectively. Gas-phase species
are on the left panels and ice mantle species on the right. Only the
region inside $2\times 10^4$ AU is shown, because densities at larger
radii are not high enough to be distinguished from the ambient gas in
observations.

\begin{figure}
\plotone{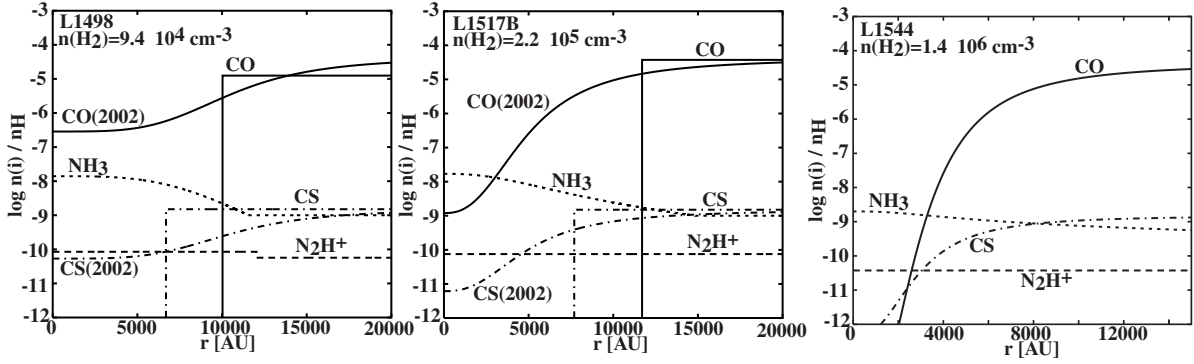}
\caption{Estimated molecular distributions in L1498, L1517B, and L1544
by Tafalla et al. (2002; 2004). Abundances of CO and CS estimated by
Tafalla et al. (2002) are labeled by CO(2002) and CS(2002). The estimated
central density $n$(H$_2$) of the core is given in each plot.
\label{Tafalla}}
\end{figure}

The molecular abundance distributions in L1498, L1517B, L1521E, L1689B,
and L1544 have been estimated by \citet{taf2002}, \citet{les2003},
\citet{taf2004}, and \citet{ts2004} based on the intensity map of dust
continuum and molecular lines. The distributions in L1498, L1517B and
L1544 are shown in Figure \ref{Tafalla} for a comparison with our model
results. Note that the five objects have different central densities:
$n (\rm H_2)=9.4\times 10^4$ cm$^{-3}$ in L1498,
$2.2\times 10^5$ cm$^{-3}$ in L1517B, $2.7\times 10^5$ cm$^{-3}$ in L1521E,
$1\times 10^6$ cm$^{-3}$ in L1689B, and $1.4\times 10^6$ cm$^{-3}$
in L1544, although these values should be taken with caution because
they depend on the somewhat uncertain core temperature. Indeed, if the center
of L1544 is as cold as 7 K, for example, the central
density increases to $\sim 10^7$ cm$^{-3}$ \citep{ev2001,zwg2001}.

It is clear that CO and CS are significantly depleted at the centers of
the three cores shown in Figure \ref{Tafalla}. But the extent of
depletion is not so clear because the observations are not
sensitive to very low abundances at the center.   For example, we cannot tell
the real CO distribution within $\lesssim 7000$ AU from the observations of
CO line intensities. \citet{taf2004} observed L1517B and L1498
with higher angular resolution than \citet{taf2002}, and found that  their
data are consistent with a step function distribution, as can be seen
in Figure \ref{Tafalla}. The degree of depletion of CO and CS in the
central regions should therefore be taken with caution.
Contrary to the three sources shown in Figure \ref{Tafalla}, no  CO
depletion is found in L1521E \citep{ts2004} and little depletion is 
found in L1689B \citep{les2003}. So, a successful dynamic/chemical
model must be able to explain this duality of results concerning
depletion.

Below we discuss our calculated distributions of individual species in some
detail, grouped by class of molecules.  We first compare radial
distributions of molecular abundances obtained in our models with the
observations. Comparison via integrated molecular column densities is
discussed in a later subsection.

\subsubsection{C-bearing species}
As expected from previous theoretical work \citep[e.g.][]{aik2003},
the central depletion of gaseous CO and other C-bearing species is more
significant at later stages of contraction and in the model of slower
contraction ($\alpha=1.1$).  At higher densities, the time scale for
adsorption onto grains is shortened, while a slower contraction gives
more time for adsorption and gas-phase destructive reactions (especially
for so-called `early-time' species) to be effective. The distributions
of gas-phase species in the model with $\alpha=4.0$ are similar to those
in the L-P model of \citet{aik2003}, which is reasonable considering
the similarity of the dynamics (\S 2).

The distributions of CO and CS in our $\alpha=1.1$ model are in
reasonable agreement with those in L1517B and L1544 at the appropriate
central core densities. Although a more detailed comparison via intensity
profiles would be desirable considering the uncertainties in the observational
estimates of their central abundances, the model with $\alpha=1.1$ is clearly
in better agreement with these cores than the $\alpha=4.0$ model.
On the other hand the shallower distributions
in L1521E are similar to the $\alpha=4.0$ case \citep{ts2004}. Note that
L1521E and L1517B have similar central densities.

Several interpretations are possible for this difference in the molecular
abundances. If contraction is ignored, as in pseudo-time-dependent models,
the degree of CO depletion at the core center is a function of time.
The adsorption time scale is given by $[\pi a^2 \sqrt{\frac{8 k T}{\pi m}}
n({\rm grain})]^{-1} \sim 2\times 10^4 (\frac{n_{\rm H}}{3\times 10^5
{\rm cm}^{-3}})$ yr, where $a$ is the grain radius (0.1 $\mu$m), $k$ is
the Boltzmann constant, $T$ is the temperature (10 K) and $m$ is the 
mass of the CO molecule. Hence the central
region of L1517B should be older and that of L1521E should be younger than
a few $10^4$ yr, which is the most naive interpretation.
However the density structures of the cores are not constant in reality;
since the central densities of these prestellar cores are high ($\gtrsim 10^5$
cm$^{-3}$), they have been formed from the ambient
low-density ($10^3-10^4$ cm$^{-3}$) component by contraction.
Given the similar central densities, the distribution of molecular abundances
should also be similar, if the contraction rate is similar in both sources.
It is then likely that any difference in the central depletion should 
probe the difference in contraction, and hence the ratio of the gravitational
force to the pressure force. There is a caveat to this line of 
reasoning, however: if the central region of L1517B is colder than 
that of L1521E, then the former may indeed be denser and its central 
depletions higher.  On the other hand, unless the central density of 
L1521E is much less than 10$^{5}$ cm$^{-3}$, the absence of 
significant CO depletion puts a rather strong constraint on its 
dynamical age.



\subsubsection{N-bearing species}
\begin{figure}
\plotone{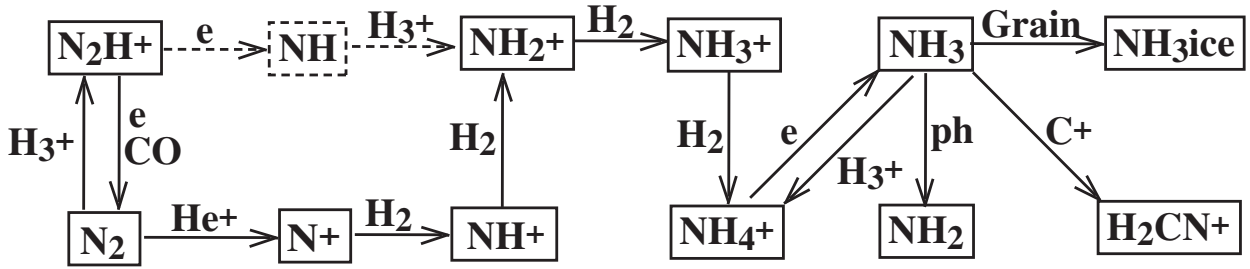}
\caption{Main formation and destruction reactions of N$_2$H$^+$ and NH$_3$
in the gas phase.
If CO is depleted, the reactions shown with dashed lines become dominant.
\label{path}}
\end{figure}

Because of the low adsorption energy of their precursor molecule N$_2$
on grain surfaces, the depletion of NH$_3$ and N$_2$H$^+$ is less significant
than that for other heavy-element species \citep{bl1997, aik2001, aik2003}.
Let us first discuss a more detailed feature: the ratio of NH$_3$ to
N$_2$H$^+$.
For L1517B and L1498, \citet{taf2002, taf2004} found the N$_2$H$^+$
abundance to be almost constant
but NH$_3$ to increase towards the center by a factor of $\sim 15$.
A smaller enhancement of $\sim 3$ is also observed in L1544 \citep{taf2002}.
Since both species are formed mainly from N$_2$ and since the molecular ion,
N$_2$H$^+$, is not considered to
be adsorbed onto grains directly, the central enhancement of NH$_3$/N$_2$H$^+$
seems puzzling, and  previous models have not shown such an enhancement.
In our $\alpha=1.1$ model, on the other hand, the NH$_3$/N$_2$H$^+$ ratio
indeed increases inwards when the central density is $n_{\rm H}=3\times 10^5$
cm$^{-3}$, which is similar to the value in L1517B and L1498. An analogous
enhancement towards the inner radius can also be seen in the region of
2500 AU $\lesssim r \lesssim 10000$ AU in our models with $\alpha=1.1$ and
4.0 when the central density is $3\times 10^6$ cm$^{-3}$, a density similar to
the value in L1544. The enhancement in our models is caused by the newly
adopted branching ratio
for the N$_2$H$^+$ recombination, which is discussed in \S 2.2.
At large radii, where gas-phase CO is abundant, N$_2$H$^+$ is destroyed mainly
by proton transfer to CO. The product, N$_2$, reacts with H$_3^+$ to reform
N$_2$H$^+$, but in the central regions, where CO is depleted, N$_2$H$^+$ mainly
recombines to produce NH, which is then transformed to NH$_3$ by reactions with
H$_3^+$ and H$_2$.  The chemical pathways are detailed in Figure \ref{path}.
The rate coefficient of N$_2$H$^+$ + CO is $1.5\times 10^{-9}$ cm$^3$
s$^{-1}$, while that of N$_2$H$^+$ + e $\to$ NH + N is $2.4 \times 10^{-6}$
cm$^3$ s$^{-1}$. Hence NH$_3$ enhancement occurs when the abundance ratio
n(CO)/n(e) becomes smaller than $\sim 10^3$.

Now let us consider the individual radial distributions of
protonated nitrogen and ammonia for $\alpha$=1.1.  As can be seen in
Figure \ref{dist_abun_alp1}, at a central density of $3\times 10^5$
cm$^{-3}$, both N$_2$H$^+$ and NH$_3$ are {\it not} seriously depleted towards
the center.  When the central density is raised to $3\times 10^6$
cm$^{-3}$, however, the species are depleted at $r\lesssim 2500$ AU.
This result contradicts the observation of
\citet{taf2002}, who found their abundances to be almost constant (with a
slightly enhanced NH$_3$/N$_2$H$^+$ ratio inwards) in L1544, which possesses a
central core density in excess of 10$^{6}$ cm$^{-3}$. It should be noted,
however, that a detailed analysis by other researchers suggests that
N$_2$H$^+$ and NH$_{3}$ could be depleted at the center
of this source \citep{cas2003}.  \citet{ba2004} also found that 
N$_2$H$^+$ is depleted in
the center of the cold protostellar core IRAM 04191, where the density
$n_{\rm H}$ is $\gtrsim 10^6$ cm$^{-3}$.

In our model, the depletion of NH$_3$ and N$_2$H$^+$ is caused by a
combination of adsorption of N$_2$ and N onto grains and transformation
of N atoms to NH$_3$ on grain surfaces, where it remains given its 
high binding energy of 3080 K.
Even with our chosen low adsorption energy for N$_2$ of 750 K, the 
rate of adsorption
for this species is larger than the cosmic-ray induced desorption rate when
the density $n_{\rm H}$ is $\gtrsim 10^6$ cm$^{-3}$.
However, the rate of cosmic-ray induced desorption is quite sensitive
to the choice of adsorption energy. If we adopt an even smaller adsorption
energy for N$_2$, and if hydrogenation of N atom is less efficient
(see \S 3.2.3 for a discussion on grain-surface reactions), the depletion
of NH$_3$ and N$_2$H$^+$ will not occur until later stages of contraction,
when the higher density leads to a higher rate of adsorption.

Although we reproduce the central enhancement of the NH$_3$/N$_2$H$^+$ ratio,
the relevant central NH$_3$ abundances calculated in our $\alpha=1.1$
model are higher than the observed values shown in Figure
\ref{Tafalla} by around an order of magnitude. On the other hand,
the calculated abundance of N$_2$H$^+$ is in reasonable agreement
with the results for all three cores shown in Figure
\ref{Tafalla}. To change the abundance of ammonia while maintaining
the same abundance of N$_{2}$H$^{+}$, one should not change the
abundance of the molecular nitrogen mother molecule or the cosmic ray
ionization rate since these parameters will affect both 
species. Ammonia in the gas phase is formed from N$^+$,
which is in turn formed by the cosmic-ray ionization of N atoms in early
stages of contraction and by N$_2$ + He$^+$ in later stages
(Figure \ref{path}). If the rate coefficients for processes leading to
N$^+$ formation are smaller than assumed in our model, the NH$_3$ abundance
become smaller, without affecting N$_2$H$^+$.

\subsubsection{Ice Composition}

The abundances of major species in ice mantles are shown in the right
panels of Figure \ref{dist_abun_alp1} and Figure \ref{dist_abun_alp4}.
With some important exceptions, it can be seen that the abundances
do not vary much with radius, especially compared with the radial
dependence of heavy gas-phase species. In addition,
the dependence on time, or central density, is not significant
when the central density
is higher than $3 \times 10^5$ cm$^{-3}$.  Mantle species are formed
both by direct adsorption from the gas and by grain-surface reactions.
As an example of surface formation,
adsorbed CO is hydrogenated to form HCO, which is further transformed
to H$_2$CO, and CH$_3$OH on grain surfaces by reactions with
atomic hydrogen. The peak abundance of CH$_3$OH in the gas-phase is
$7\times 10^{-12}$, which is much lower than its ice abundances. Carbon
dioxide in ice mantles is formed mainly by a surface reaction between O and
HCO on grain surfaces. In our model, the activation barrier of this reaction is
assumed to be zero. Hence it is more efficient than another CO$_2$-forming
reaction, CO + O, for which we have assumed an activation barrier of 1000 K.
This older value is somewhat higher than the value of 290 K 
measured by \citet{ros01} on amorphous ice. Use of this lower value 
at 10 K does not result in significant changes to our results.

The ice composition in dense prestellar cores has not been observed so far,
because of the faintness of background stars.
Instead, we compare our model results with observations towards low-mass
protostars and field stars.
Although the central densities of our model cores are very high ($\gtrsim
10^5$ cm$^{-3}$), the density at $r\sim 20000$ AU is as low as $n_{\rm H}\sim
6\times 10^3$ cm$^{-3}$, which is comparable to that of the foreground
material in front of these stars. So we can compare our surface
predictions for this large radius. It is customary to describe abundances of
various surface species relative to H$_2$O ice.  With this convention,
Table \ref{ice} summarizes ice compositions observed towards the 
field star Elias 16, the low-mass protostar Elias 29, and the high-mass
protostars NGC 7538 IRS9, GL 7009S and W33A \citep{es2000,gib2000}.
It can be seen that carbon monoxide and carbon dioxide ices are 
relatively abundant toward these sources. The abundance of CH$_3$OH is high
towards high-mass protostars, while it is low towards the field star
Elias 16 and the low-mass protostar Elias 29. It should be noted, however,
that solid CH$_3$OH has recently been detected
with an abundance of $15-25$ \% towards three low-mass YSO's \citep{ppp2003}.

\begin{deluxetable}{l r r r r r r r r}
\tabletypesize{\scriptsize}
\tablecaption{Ice Composition in Molecular Clouds and Protostellar Cores
\tablenotemark{a}
\label{ice}}
\tablewidth{0pt}
\tablehead{
\colhead{Species} &\colhead{field star} &
\colhead{low-mass PS\tablenotemark{b}} &
\colhead{} & \colhead{high-mass PS} & \colhead{} & \colhead{} &
\colhead{model} & \colhead{}\\
\colhead{} & \colhead{Elias 16} & \colhead{Elias 29} &
\colhead{NGC 7538 IRS9} & \colhead{GL 7009S} & \colhead{W33A} &
\colhead{10 K} & \colhead{12 K} & \colhead{15 K}}

\startdata
H$_2$O  &  100 &    100 & 100 & 100 & 100 & 100 & 100 & 100\\
CO      &   34 &    5.6 &  16 &  15 &   8 &  29 & 0.01& 0.4\\
CO$_2$  &   15 &     22 &  20 &  21 &  13 &  0.6& 0.5 &  68\\
CH$_4$  &   -  & $<1.6$ &   2 &   4 & 1.5 &  3.3& 5.6 & 5.8\\
CH$_3$OH&$<3.4$& $ <4 $ &   5 &  30 &  18 &  5.8& 37  & 9.5 \\
H$_2$CO &   -  &    -   &   2 &   3 &   6 & $10^{-4}$& 0.01& 0.4\\
NH$_3$  &$<6$  & $<9.2$ &  13 &   - &  15 & 0.4 & 17  &  18\\
\enddata


\tablenotetext{a}{\citet{es2000}; \citet{gib2000}}
\tablenotetext{b}{PS stands for protostar.}


\end{deluxetable}

\begin{figure}
\plotone{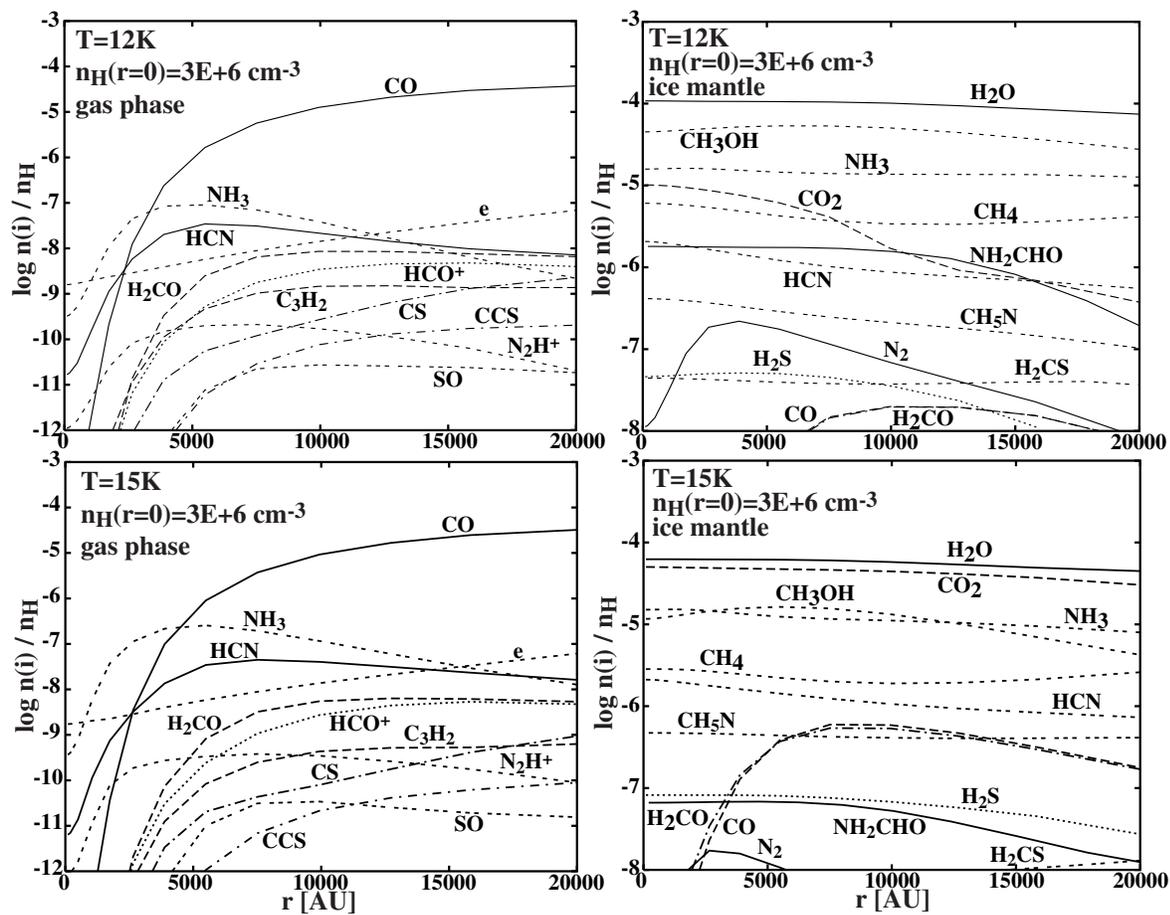}
\caption{Distributions of molecular abundances as in Figure
\ref{dist_abun_alp1} but for $T=12$ K and 15 K.
The central density of the core is $3\times 10^6$ cm$^{-3}$.
\label{dist_T12}}
\end{figure}

Listed in the right three columns of Table \ref{ice} are ice compositions
at $r=20000$ AU in the model with $\alpha=1.1$ and central density of
$3\times 10^6$ cm$^{-3}$. With our 10 K model, the  density at 
this radius remains slightly under 10$^{4}$ cm$^{-3}$. We varied 
the temperature of the core, since portions of the gas in front of
the sources considered here may well be warmer than 10 K. Although
the contraction time scale depends on the sound velocity within the core,
the dynamics have been assumed to be the same as the 10 K model with
$\alpha=1.1$ in order to probe only the chemical effects.
Distributions of molecular abundances for models with $T=12$ K and 15 K
are shown in Figure \ref{dist_T12} . The predicted ice composition depends
sensitively on the temperature, because of the low adsorption and migration
energies of reactant species such as H atoms.

We first compare the $T=10$ K model
with observations towards Elias 16, which correspond to the ice composition
in a cold quiescent cloud. The calculated abundances of CO and NH$_3$ 
are consistent, but the
CH$_3$OH abundance is higher and the CO$_2$ abundance significantly lower than
the observed values. The model with $T=15$ K, on the other hand,
gives a high CO$_2$ abundance and little CO. Since there are temperature
variations within a molecular cloud, a mixture of $T=10- 15$ K components
may reproduce the observed CO and CO$_2$ abundances better. However, 
such a mixture would give too much CH$_3$OH. The material in front of Elias 16 
may well have lower gas densities than considered here; lower-density 
results toward the edge of our model core indeed show lower methanol 
abundances. Next we compare our models with low- and
high-mass protostars. Both a warm envelope and cold foreground
component exist along the observed line of sight so that, on average,
the temperature of the matter along the line of sight towards these sources
may be higher than that towards Elias 16. The calculated abundances 
of NH$_3$ and
CH$_3$OH ices in our models with $T=12$ and 15 K are consistent with
the observations, and a mixture of these two models can reproduce the
CO$_2$ abundance. However, the CO and H$_2$CO abundances are significantly
underestimated.

Although the possible temperature variation along the observed line of sight
makes the comparison difficult, it seems that our model generally 
underestimates CO
at $T>10$ K, and CO$_2$ and H$_2$CO at $T<15$ K. In the models with $T=12$
and 15 K, CO ice is transformed to CH$_3$OH and CO$_2$ by grain-surface
reactions in manner that seems to be too efficient. Destruction of CO would be
suppressed if we take into account the recent experimental results that
CO hydrogenation occurs only within several surface layers, because H atoms
cannot penetrate deep into ice mantles \citep[][Chigai et al. in prep]{hid2004,
wnsk2004}.  Such an effect can be accounted for with the 
so-called three-phase approach of \citet{hh1993b}. In order to 
improve the model, it would also be useful to calibrate
the rate-equation formula for the grain surface reactions. Although the
ice composition calculated by the modified-rate method is in good agreement
with that by stochastic methods \citep{cas2002b,sh2004}, a small
grain-surface-reaction network (i.e. including less than a few tens of
reactions) is used in such comparisons. On the other hand, our current model
includes more than 500 grain-surface reactions (with deuterium species).
Calibration of the modified-rate method with a stochastic method
using a large reaction network is desirable.

\subsection{Molecular Column Densities}

\begin{deluxetable}{l c c c c c c c}
\tabletypesize{\scriptsize}
\tablecaption{Molecular Column Densities (cm$^{-2}$) of Gaseous
Species in Theoretical Models
\label{column}}
\tablewidth{0pt}
\tablehead{
\colhead{Species} &\colhead{} & \colhead{alpha=1.1} & \colhead{} & \colhead{}
& \colhead{} & \colhead{alpha=4.0} & \colhead{}\\
\colhead{} & \colhead{$n_{\rm H}=3\times 10^5$ cm$^{-3}$} & \colhead{$3\times
10^6$ cm$^{-3}$} & \colhead{$3\times 10^7$ cm$^{-3}$} & \colhead{} &
\colhead{$3\times 10^5$ cm$^{-3}$} & \colhead{$3\times 10^6$ cm$^{-3}$} &
\colhead{$3\times 10^7$ cm$^{-3}$}}

\startdata
CO          & 1.9(17)\tablenotemark{a} & 1.9(17) & 2.0(17) & &
1.2(18) & 1.0(18) & 1.0(18) \\
C$_3$H$_2$  & 9.8(12) & 8.7(12) & 8.8(12) & & 2.2(14) & 1.6(14) & 1.5(14) \\
H$_2$CO     & 5.1(13) & 4.6(13) & 4.6(13) & & 1.1(15) & 7.9(14) & 7.3(14) \\
CH$_3$OH    & 5.9(10) & 5.3(10) & 5.3(10) & & 9.6(12) & 4.3(12) & 3.5(12) \\
CCS         & 7.5(11) & 6.9(11) & 6.9(11) & & 2.5(13) & 1.8(13) & 1.7(13) \\
CS          & 1.1(13) & 1.0(13) & 1.0(13) & & 6.6(13) & 5.4(13) & 5.2(13) \\
SO          & 3.2(11) & 3.5(11) & 3.9(11) & & 1.6(13) & 1.5(13) & 1.4(13) \\
HCN         & 1.0(15) & 8.0(14) & 7.8(14) & & 5.5(14) & 7.9(14) & 7.3(14) \\
HC$_3$N     & 7.4(12) & 5.8(12) & 5.8(12) & & 5.1(13) & 4.4(13) & 4.1(13) \\
NH$_3$      & 6.2(15) & 7.8(15) & 7.4(15) & & 7.0(14) & 4.8(15) & 5.4(15) \\
N$_2$H$^+$  & 1.1(13) & 1.8(13) & 2.0(13) & & 5.5(12) & 2.6(13) & 3.5(13) \\
N$_2$D$^+$  & 1.7(12) & 4.2(12) & 5.2(12) & & 5.5(11) & 6.7(12) & 1.1(13) \\
HCO$^+$     & 2.9(13) & 2.9(13) & 3.0(13) & & 1.4(14) & 1.3(14) & 1.3(14) \\
DCO$^+$     & 2.3(12) & 2.4(12) & 2.6(12) & & 1.0(13) & 1.4(13) & 1.3(13) \\
H$_3^+$     & 5.3(13) & 6.3(13) & 5.9(13) & & 3.9(13) & 5.3(13) & 5.4(13) \\
H$_2$D$^+$  & 2.3(13) & 6.3(13) & 7.9(13) & & 9.1(12) & 4.3(13) & 7.4(13) \\
\enddata


\tablenotetext{a}{$a(b)$ means $a\times 10^b$.}


\end{deluxetable}

In order to derive molecular radial distributions from the observational data,
one has to solve a (non-LTE) radiation transfer problem with some
assumptions such as
the density and velocity distributions. For easier and more straightforward
comparison between observational data and theoretical models, it is useful
to calculate molecular column densities integrated along the line of sight
in the models. Table \ref{column} lists column densities of assorted
gas-phase molecular species in our models with $\alpha=1.1$ and 4.0
when the central
density is $3\times 10^5$, $3\times 10^6$, and $3\times 10^7$ cm$^{-3}$.
We first integrate the absolute molecular abundance (i.e. number density)
along the line of sight to obtain a
column density at each impact parameter, and then average them within the
beam size of the IRAM 30 m telescope, except for NH$_3$, H$_3^+$ and
H$_2$D$^+$.
The column density of NH$_3$ is averaged over a radius of 20'', and H$_3^+$
and H$_2$D$^+$ are averaged over a radius of 11'', considering the beam
sizes of the Effelsberg and CSO telescopes, respectively.

As can be seen in Table \ref{column}, the column densities of
carbon-bearing species do not vary with time while
the central density increases from $3\times 10^5$ cm$^{-3}$ to $3\times 10^7$
cm$^{-3}$. Although the central depletion becomes heavier as contraction
proceeds, contributions from the outer radii keep the molecular column
densities almost constant. In order to estimate the evolutionary stage
(i.e. central density) of a core from these species, a comparison with the
radial abundance distribution is more useful. On the other hand, many of the
column densities have a clear dependence on
$\alpha$; they are higher in the model of larger $\alpha$, where the
contraction is faster.

As opposed to the behavior of carbon-bearing species, the column
densities of nitrogen-bearing and deuterated species increase
as the central density rises. This dependence is more significant in the model
with larger $\alpha$.
The depletion of N$_2$H$^+$ and NH$_3$ becomes
significant only inside $r\sim 2500$ AU and in a late stage of contraction,
because of the low adsorption energy of their mother molecule N$_2$.
As the central density increases from $3\times 10^5$ to $3\times 10^7$
cm$^{-3}$, the radial distributions of the N$_2$H$^+$ and NH$_3$ abundances
relative to hydrogen do not change much in the region of $2500$ AU $\lesssim
r\lesssim 5000$ AU (Fig. \ref{dist_abun_alp1} and Fig. \ref{dist_abun_alp4}),
which means that their absolute abundances there increase as the core
contracts.

\begin{deluxetable}{l c c c c c c}
\tabletypesize{\scriptsize}
\tablecaption{Observed Molecular Column Densities in Assorted  Prestellar Cores
\label{survey}}
\tablewidth{0pt}
\tablehead{
\colhead{Object} &\colhead{central density} & \colhead{N$_2$H$^+$}   &
\colhead{NH$_3$} & \colhead{CCS}  & \colhead{CS} & \colhead{CO} \\
\colhead{}&\colhead{$n$(H$_2$) cm$^{-3}$} & \colhead{$10^{12}$ cm$^{-2}$} &
\colhead{$10^{14}$ cm$^{-2}$} & \colhead{$10^{12}$ cm$^{-2}$} &
\colhead{$10^{14}$ cm$^{-2}$} & \colhead{$10^{17}$ cm$^{-2}$}\\}

\startdata
L1498   & $9\times 10^4$ & 8$\pm$ 4\tablenotemark{a}, 3.0\tablenotemark{b} &
  (4.1)\tablenotemark{c,d}, 2.3\tablenotemark{b} &
  (16.5)\tablenotemark{c} & (0.40)\tablenotemark{e}, 0.14\tablenotemark{b} &
   2.4\tablenotemark{b} \\
L1517B  & $2.2\times 10^5$ & 3$\pm 0.3$\tablenotemark{a}, 3.1\tablenotemark{b}&
   7\tablenotemark{c}, 2.1\tablenotemark{b} &
   8.6\tablenotemark{c} & 0.13\tablenotemark{b} & 1.5\tablenotemark{b}  \\
L1521E  & $2.7\times 10^5$ & $<1.4$\tablenotemark{f} & 0.73\tablenotemark{f} &
   28\tablenotemark{f} & 3.0\tablenotemark{f} & 8.5\tablenotemark{f}\\
L1544   & $1.4\times 10^6$ & 9$\pm 2$\tablenotemark{a}, 7.3\tablenotemark{b} &
   1.8\tablenotemark{b} &
   20\tablenotemark{g} & 0.46\tablenotemark{b} & 4.0\tablenotemark{b},
   12\tablenotemark{h} \\
\enddata


\tablenotetext{a}{\citet{cas2002d}}
\tablenotetext{b}{\citet{taf2002}}
\tablenotetext{c}{\citet{su92}}
\tablenotetext{d}{The column densities with parentheses are observed
somewhat ($\sim 30''$) offset from the N$_2$H$^+$ peak.}
\tablenotetext{e}{\citet{hir1998}, assuming $^{34}$S/S= 4.2\%}
\tablenotetext{f}{\citet{hir2002} and references therein}
\tablenotetext{g}{\citet{oha1999}}
\tablenotetext{h}{\citet{cas2002a}}


\end{deluxetable}

Table \ref{survey} shows estimated central densities and determined molecular
column densities in some prestellar cores.
The column density of N$_2$H$^+$ is in agreement with our models ($\alpha=1.1$
and 4.0) within a factor of a few, except for L1521E, which has a very
low N$_2$H$^+$ column density in spite of its relatively high central
density. If the central density is of order $10^5$ cm$^{-3}$, the larger
$\alpha$ is preferred for L1521E.  The ammonia abundance, and thus its column
density, are mostly overestimated in our models, as discussed in the previous
subsection. The column density of CCS is better reproduced
in the model with $\alpha=4.0$. The $\alpha=1.1$ model better reproduces
the CS column density in L1498 and L1517B, while $\alpha=4.0$ is
preferred for CS in
L1544. However, comparison of these S-bearing species should be taken with
caution, because their abundances depend on the assumed elemental abundance
of sulfur and chemical network (e.g. NSM or UMIST) \citep{aik2003}.
The observed CO column densities are in reasonable agreement with our 
model results when $\alpha=1.1$.

\subsection{Deuterium Fractionation}

\subsubsection{Mono-deuterated Species}
\begin{figure}
\plotone{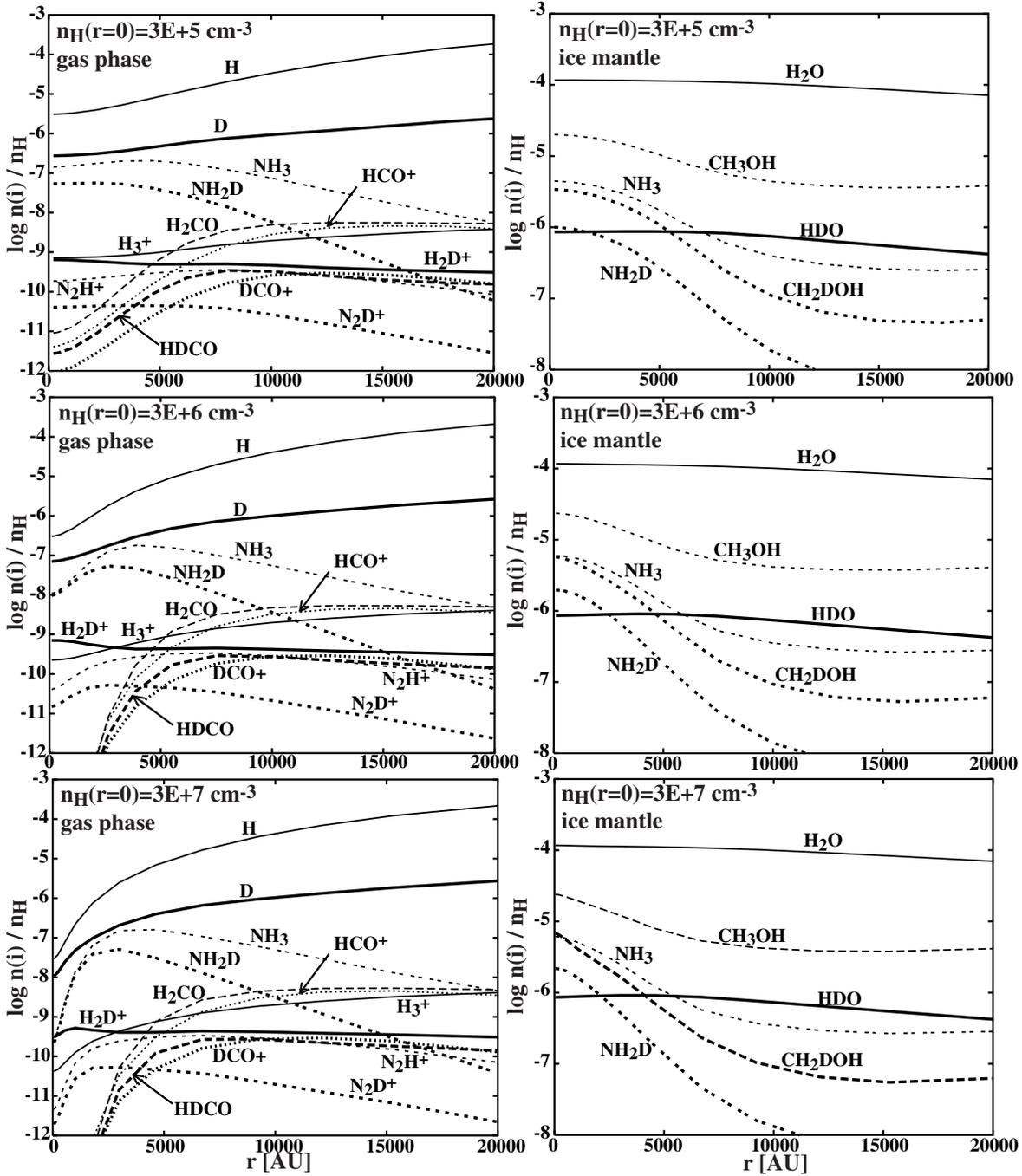}
\caption{Distribution of mono-deuterated species and their normal counterparts
with $\alpha=1.1$. Deuterated species are shown with thick lines while
their normal counterparts are shown with thin lines. Isotopomers of the same
chemical species are presented by the same type of lines (solid, dashed or
dotted). Other details are the same as in Figure 2.
\label{dist_monoD_alp1}}
\end{figure}

\begin{figure}
\plotone{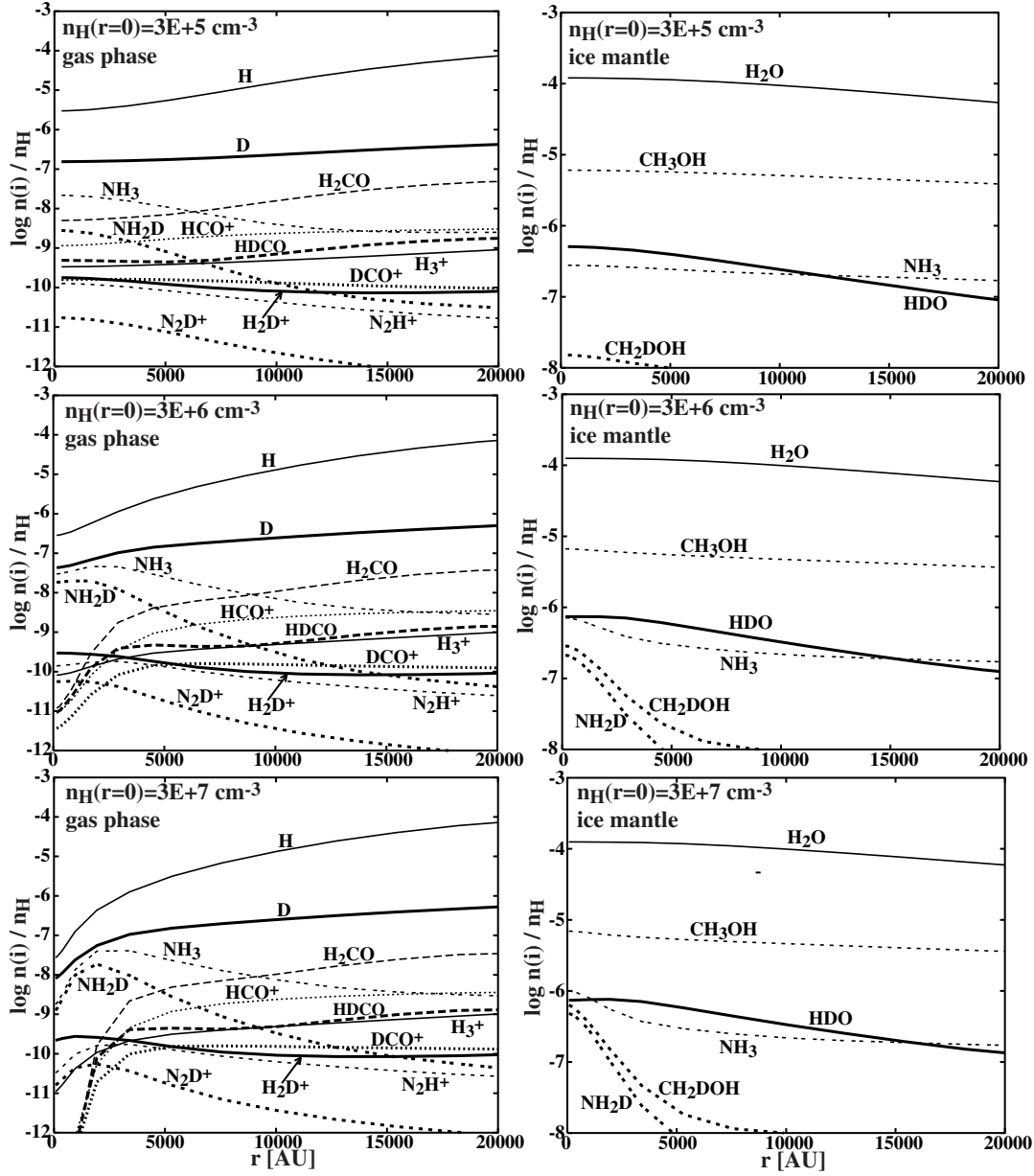}
\caption{Distribution of mono-deuterated species and their normal counterparts
as in Figure \ref{dist_monoD_alp1} but for  $\alpha=4.0$.
\label{dist_monoD_alp4}}
\end{figure}

Mono-deuterated species and deuterium fractionation are included both in
gas-phase and grain-surface reactions in our chemical reaction network
\citep[][and references therein]{aik2003}. In low-temperature cores, molecular
D/H ratios are enhanced by exothermic exchange reactions such as
H$_3^+$ + HD $\to$ H$_2$D$^+$ + H$_2$, in which the backwards endothermic
reactions are too slow to be of importance.  Since the main reactant
of H$_2$D$^+$
is CO, the D/H ratios are further enhanced by  CO depletion.
Although the back reaction between H$_2$D$^+$ + H$_2$ can be speeded
up by the presence of ortho-H$_{2}$ \citep{ger2002, wff2004}, this
process is not included in our model.
If ortho-H$_2$ is abundant, molecular D/H ratios
would become smaller than predicted here.
Computed spatial distributions of mono-deuterated
species and their normal counterparts are shown in Figure \ref{dist_monoD_alp1}
and Figure \ref{dist_monoD_alp4} for models with $\alpha= 1.1$ and 4.0,
respectively. Molecular D/H ratios are
higher in the model with smaller $\alpha$, which shows a heavier CO depletion.
The difference is especially significant
at an earlier stage of contraction, with a lower central density.

Let us now consider some important D/H ratios.  For
$\alpha=1.1$, the atomic D/H ratio in the central region
is as high as 0.23 and 0.37 when the central density is $3\times 10^6$
cm$^{-3}$ and $3\times 10^7$ cm$^{-3}$, respectively.
The high abundance of D atoms enhances the deuteration of ice-mantle species
in spite of the slower migration of D atoms compared with that of H
atoms on  grain surfaces.
The D/H ratio for the molecular ion H$_2$D$^+$ in the gas phase is higher
than unity at the core center when the central density is $\gtrsim 10^6$
cm$^{-3}$ for both values of $\alpha$.
The H$_2$D$^+$ fractional abundance of $\sim 10^{-9}$ at the
appropriate central core
density obtained in our model with $\alpha=1.1$ is consistent with the
observation in L1544 \citep{cas2003}. While \citet{cas2003} and \citet{vdt2004}
estimate that the H$_2$D$^+$ fractional abundance at the 20'' 
off-set position is smaller
than the central value by a factor of $2 - 5$, its abundance at $r\sim 3000$
AU (which corresponds to 20'' assuming a distance of 150 pc) is 
calculated to be smaller than
the central value by a factor of 1.5 when the central density is $3\times
10^6$ cm$^{-3}$.
Note that the calculated fractional abundance of H$_2$D$^+$ is relatively high
even in the latest contraction stage, when the central density is
$3 \times10^7$ cm$^{-3}$ in our model, but will promptly decrease
when the gas is heated by a newly born protostar, because the endothermic
backwards reaction between H$_{2}$D$^{+}$ and H$_{2}$ will begin to turn on.
Hence we conclude that H$_2$D$^+$ is a good probe of starless cores immediately
prior to star formation.

How do our results compare with observation of other singly deuterated species?
\citet{cas2002a} obtained column density ratios for
$N({\rm DCO}^+)/N({\rm HCO}^+)$ and $N({\rm N}_2{\rm D}^+)/
N({\rm N}_2{\rm H}^+)$ towards the dust peak of L1544 of 0.04 and
0.2, respectively.
In our model with $\alpha=1.1$ and a central density of $3\times
10^6$ cm$^{-3}$, the calculated column density ratios are $N({\rm DCO}^+)/
N({\rm HCO}^+)=0.08$ and $N({\rm N}_2{\rm D}^+)/N({\rm N}_2{\rm H}^+)=0.2$
(Table \ref{column}), both in reasonable agreement with the observed
values.  These results for singly-deuterated species can be changed
when multiple deuteration is considered; the calculated D/H ratios 
for DCO$^+$ and N$_2$D$^+$, however, are larger by only a factor of 1.5
in the multi-deuterated model discussed below.

\subsubsection{Multi-deuterated Species}

In recent years, multi-deuterated species have been detected in
several prestellar
and protostellar cores. For example, \citet{vdt2002} and \citet{lis2002}
detected ND$_3$ and derived relative abundances of $n$(ND$_3$)/$n$(H$_2$)
$\sim 10^{-12}-10^{-11}$ and $n$(ND$_3$)/$n$(NH$_3$) $\sim 10^{-3}$
in NGC 1333 and Barnard 1. \citet{bac2003} obtained $n$(D$_2$CO)/$n$(H$_2$)
$\sim 10^{-11}$ and $n$(D$_2$CO)/$n$(H$_2$CO)$\sim 0.01-0.1$ in several
prestellar cores.
In addition, the mono-, doubly-, and triply-deuterated methanol
isotopomers CH$_{2}$DOH, CHD$_{2}$OH and CD$_3$OH have abundances of
$0.3\pm 0.05$, $0.06\pm 0.01$, and $0.014 \pm 0.006$ relative to their
normal counterpart, respectively, in the protostar IRAS 16293
\citep[][B. Parise personal communication]{par2004}.
These highly-deuterated methanol isotopomers  are
considered to be formed
by grain-surface reactions in the prestellar stage and desorbed when the
core is heated by a newly born star. As shown in the previous
subsection, our model produces high abundances of mono-deuterated species,
including atomic deuterium, which is important for subsequent deuteration on
grain surfaces. Hence it is interesting to see if we can also obtain large
enhancements for multi-deuterated species.

\begin{figure}
\plotone{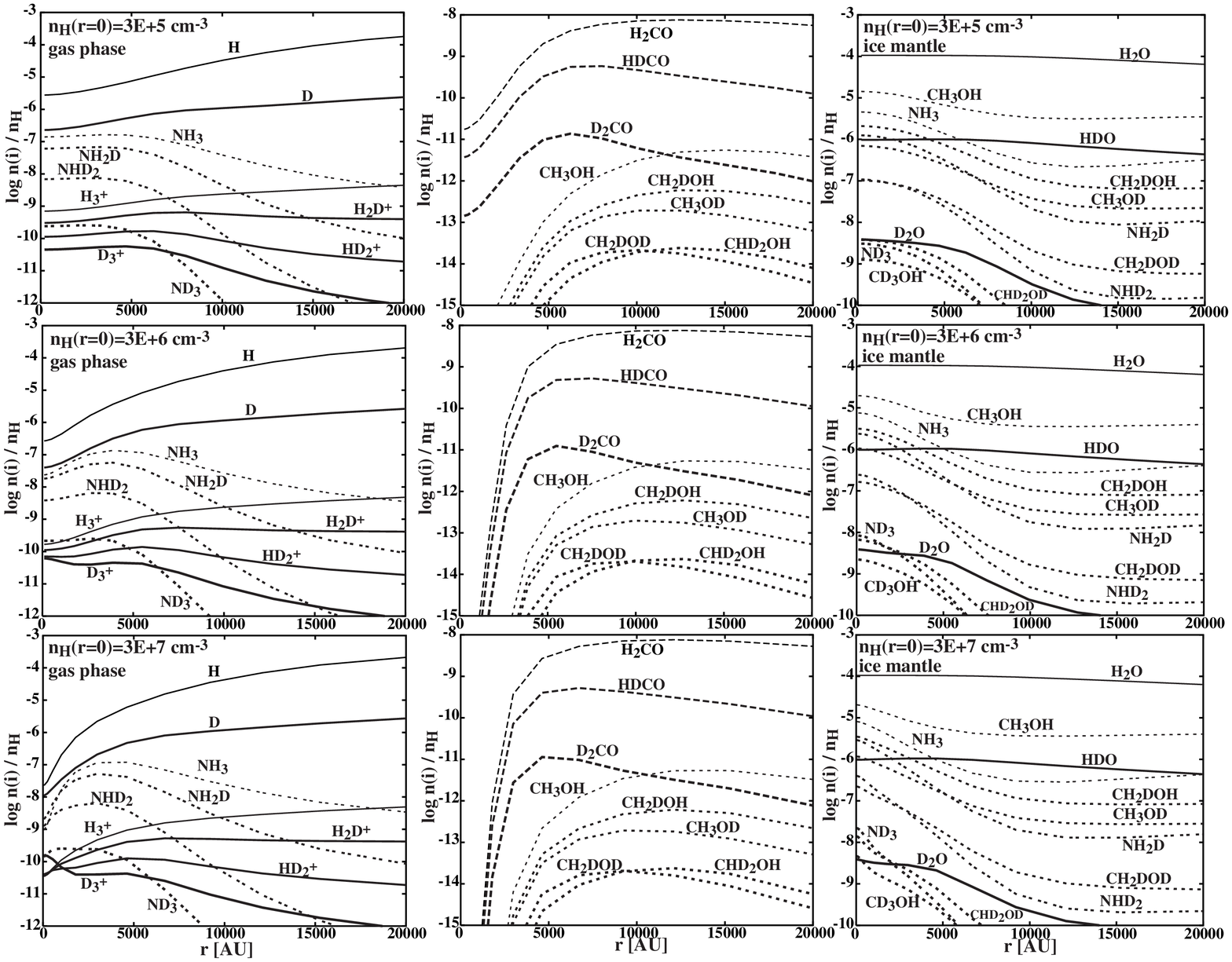}
\caption{Distribution of multi-deuterated species and their normal counterparts
with $\alpha=1.1$. Other details are the same as in Figure 7.
\label{multi_D}}
\end{figure}

The left and middle panels in Figure \ref{multi_D} show radial
distributions of assorted gas-phase species and their singly and
multiply deuterated isotopomers obtained from our $\alpha=1.1$ model.
Comparison of the calculated D/H ratio with the observed value is tricky;
while the molecular D/H ratios increase, the abundances of most gas-phase
species relative to hydrogen decrease inwards. One should calculate molecular
column densities considering the core structure, in order to compare the model
results with observations.

Let us focus attention on D$_{2}$CO and ND$_{3}$.
Since the critical densities of the observed D$_2$CO transitions
($J=2-1$ and $J=3-2$) are relatively high ($\gtrsim 10^5$ cm$^{-3}$),
the observed abundances are compared with our results in the central
regions. In particular, we calculated the column densities of H$_2$CO and
D$_2$CO in the region of $n_{\rm H}\ge 2\times 10^5$ cm$^{-3}$ by
integrating along the line of sight, and then obtained the column density
ratio, which is $\sim 7\times 10^{-3}$ throughout our range of central
densities $3\times 10^5$ cm$^{-3}- 3\times 10^7$ cm$^{-3}$, when the CO
depletion factor ($N$(H$_2$)/$N$(CO)$\times 10^{-4}$) is $12-250$.
On the other hand, \citet{bac2003} obtained D$_2$CO/H$_2$CO column density
ratio of 0.01-0.14 (including error bars) in prestellar cores with the CO
depletion factor of $\sim 15$.
Although the peak abundance of D$_2$CO in our $\alpha=1.1$ model is similar
to the observed value ($\sim 10^{-11}$) in these prestellar cores,
the D$_2$CO/H$_2$CO ratio is clearly underestimated.
Evaporation of D$_{2}$CO and H$_2$CO from ice mantles
does not improve the agreement in the gas phase, since the surface abundance
ratio for D$_2$CO/H$_2$CO is lower than
that in the gas phase. This disagreement is not found in the
stationary shell deuterium-fractionation model of \citet{rob2004},
where the calculated ratio of 0.03 for D$_{2}$CO/H$_{2}$CO is in good
agreement with that measured for L1544. The higher fractionation 
stems partially from the use of the UMIST RATE99 network of reactions 
\citep{let00,rob2004}.

NGC 1333 and Barnard 1, in which ND$_3$ is observed, harbor protostars.
Since the protostellar stage is not included in our model, we compare our
results at the latest stage, at which a  central density of $3 \times
10^7$ cm$^{-3}$ is reached, with observations.
The abundance ratio of ND$_3$ and NH$_3$ varies significantly with radius.
The beam size of the ND$_3$ observations is 25'', which corresponds to a
spatial resolution of $\sim 8800$ AU assuming that the distance to the objects
is 350 pc. Hence we calculate the column density ratio of ND$_3$ and NH$_3$
averaged over a radius of 4400 AU; the ratio is $2\times 10^{-3}$ and is in
reasonable agreement with observations. It should be noted that our
peak abundance of ND$_3$ is $\sim 10^{-10}$, which is much higher than
the observed values ($\sim 10^{-11}$). As we discussed in \S 3.2.2, our
model overestimates the abundance of NH$_3$. If we could reduce the
abundance of
N$^+$, both NH$_3$ and ND$_3$ abundances would be decreased without affecting
their abundance ratio.

The distributions of ice-mantle isotopomers are shown in the right panels
of Figure \ref{multi_D}. In the central region of the core with a central
density of $\gtrsim 3\times 10^6$ cm$^{-3}$, the abundance ratio of ND$_3$
to NH$_3$ is about $10^{-3}$. Hence, even if there is some desorption of
NH$_3$ ice in the central region of these protostars, the
gas-phase abundance ratio of ND$_3$/NH$_3$ remains as high as $10^{-3}$.

Now let us consider the deuterium fractionation of methanol in ice mantles.
Methanol is  formed mainly via grain-surface
reactions, in particular the surface hydrogenation of CO by H atoms.
In fact its gas-phase fractional abundance in our model is computed to
be at most $10^{-11}$, which is less than the
observed value of $4.4\times 10^{-9}$ in IRAS 16293 \citep{vds95}.
Hence methanol and its isotopes observed in this object are
desorbed from the grain surfaces as a result of heating by a protostar,
although the desorption might be significant only in the central regions,
because the observed CH$_3$OH abundance in the gas-phase is much lower than the
value expected from the ice composition of CH$_3$OH/H$_2$O =15-25\% recently
obtained towards a low-mass protostar \citep{ppp2003}.
In the right panels of Figure \ref{multi_D}, it can be seen that the relative
abundances of the multi-deuterated methanol isotopomers to their normal
counterpart are lower than the observed values in IRAS 16293.
For example, the abundances of surface CHD$_{2}$OH (the abundance of which
is the same as that of CH$_2$DOD depicted in Figure \ref{multi_D})
and CD$_3$OH relative to normal methanol (see Figure \ref{multi_D})
are calculated to be $\sim 10^{-2}$ and $\sim 10^{-4}$ at
the center of our densest core. These values are smaller than the 
observed values of $0.06\pm 0.01$ and $0.014\pm 0.006$, respectively,
in IRAS 16293 \citep[][B. Parise personal communication]{par2004}.
In addition, the singly deuterated methanol CH$_{2}$DOH is
calculated to be $\sim 0.17$ times the abundance of normal
methanol, which is slightly lower than the claimed ratio of $0.3\pm 0.2$ in
IRAS 16293. \citet{par2004} concluded that the high
abundances of multi-deuterated species are reproduced if the atomic D/H ratio
in the gas phase is 0.1-0.3. Although the atomic D/H ratio obtained in our
model (left panels of Figure \ref{multi_D}) is similar to this value, such
a high ratio is obtained only after the adsorption of CO onto grains. Before
the atomic D/H ratio becomes high enough, a significant amount of CH$_3$OH is
formed, which lowers the total D/H ratio in methanol.


In order to improve the agreement with the methanol isotopomeric
observations, a layered structure of the ice mantle could be important.
Because the D/H ratio increases with time and
because the desorption rate of CH$_3$OH is small, those isotopomers
formed with low D/H ratio in the early stages will be buried deep in
the ice mantle while those with high D/H ratio will be in the surface layers
when the protostar is born. Then the latter component will be desorbed first.
In fact, the rotational temperature of CH$_3$OH and its isotopes in IRAS 16293
is $\sim 50$ K, which suggests that the observed region is not hot enough to
desorb the entire ice mantles. This scenario is bolstered by 
the low observed abundance of CH$_3$OH in the gas-phase of
$4.4\times 10^{-9}$ in IRAS 16293 \citep{vds95} compared with the calculated
abundance of CH$_3$OH ice obtained in our model and the observed ice abundance
in some low-mass protostellar cores (\S 3.2.3). A test for this scenario
is the observation of D/H ratios in young hotter cores in which the ice
mantles have been completely desorbed but gas-phase chemistry has not yet
affected the desorbed species significantly.

It is also interesting to look at our predictions for the abundances
of the ions D$_2$H$^{+}$ and D$_3^+$ at the centers of cores (Figure
\ref{multi_D}). In agreement with the work of \citet{rhm2003}, we find
that the latter ion is the dominant species of the H$_{\rm n}$D$_{\rm
3-n}^{+}$ family at densities $\gtrsim 10^{7}$ cm$^{-3}$. The abundance of
D$_2$H$^+$ is $\sim 10^{-10}$, which is in agreement with the first
detection of this species in the prestellar core IRAS 16293E by
\citet{vpy2004}. On the other hand, the radial distribution of 
H$_{2}$D$^{+}$ is now predicted to be rather flat or even decreasing 
towards the center (see  Figure \ref{multi_D}) at a central density 
of 3 $\times 10^{6}$ cm$^{-3}$, in disagreement with the L1544 
observations of \citet{cas2003} and \citet{vdt2004}.  This problem 
also occurs with the static shell model of \citet{rob2004}.

\section{Discussion}

\subsection{Chemical and Kinetic Properties of Observed Prestellar Cores}

We now revisit the observed properties of the prestellar cores L1498, L1517B,
L1521E, and L1544, and discuss which model (i.e. $\alpha=1.1$ or 4.0) is
preferable for each object. The upper part of Table \ref{comp} summarizes
chemical properties that are described in \S 3.2. The values in brackets
are the preferred $\alpha$ for each property and
object.  Starting with depletion, we see that $\alpha=1.1$ is preferable for 
reproducing the heavy depletion of CO
and CS in L1498, L1517B, and L1544, while the model with $\alpha=4.0$ accounts
for no depletion in L1521E. The abundance ratio of NH$_3$/N$_2$H$^+$ is
enhanced towards the core center by a factor of $\sim 15$ in L1498 and L1517B
and by a factor of $\sim 3$ in 1544. Considering their central densities,
$\alpha=1.1$ is preferable for L1498 and L1517B, while both models
($\alpha=1.1$ and 4.0) can reproduce the modest enhancement in L1544.

Comparison via molecular column densities are not listed, since these distinctions
are
less obvious than the spatial distribution; the preferable $\alpha$ varies
with species for each object. The molecular column density is directly affected
by uncertainties in both observation (e.g.  assumed geometry, velocity field
and temperature in the analysis) and reaction rate coefficients (\S 3.3).

The kinematic properties of the objects are listed in the lower part of
Table \ref{comp}. Subtracting the thermal component from the intrinsic
line width, the non-thermal components of the N$_2$H$^+$ line width are calculated to
be 0.17 km s$^{-1}$ in L1498 and L1517B \citep{taf2002}, and 0.27 km s$^{-1}$
in L1521E. \citet{hir2002} also obtained a relatively large intrinsic line
width of $0.3-0.5$ km s$^{-1}$ for species such as C$_3$S ($J=4-3$) and
H$^{13}$CO$^+$ ($J=1-0$) in L1521E. Infall velocities in L1498 and L1544
are estimated from CS and N$_2$H$^+$ lines by \citet{lee2001}.
\citet{oha1999} observed CCS in L1544, and derived a larger infall velocity
(0.1 km s$^{-1}$) than \citet{lee2001}.
While $\alpha=1.1$ is preferable for reproducing
the narrow line width and small infall velocities in L1498, L1517B, and L1544,
the large widths may be an indication of faster contraction
(i.e. $\alpha=4.0$) in L1521E.

Combining the chemical and kinematic properties, the model with $\alpha=1.1$
is consistent with L1498, L1517B, and L1544, while the model with $\alpha=4.0$
is preferable for L1521E.
Although it is not listed in Table \ref{comp}, L1689B could be another
example which is fit better by $\alpha=4.0$. \citet{les2003}
found that the molecular depletion in L1689B is much less significant
than that of L1544, in spite of the fact that their central densities are
similar. Molecular line widths in L1689B are also broader, indicating
more active kinematics than those in L1544 \citep{les2003}.


\begin{deluxetable}{l c c c c}
\tabletypesize{\scriptsize}
\tablecaption{Properties of the Observed Prestellar Cores \label{comp}}
\tablewidth{0pt}
\tablehead{
Parameter &\colhead{L1498} & \colhead{L1517B}   &
\colhead{L1521E} & \colhead{L1544}
}

\startdata
\multicolumn{5}{c}{Chemical Properties}\\
\hline
CO depletion & heavy & heavy & no & heavy \\
             & [1.1]\tablenotemark{a} & [1.1] & [4.0] & [1.1] \\
CS depletion & heavy & heavy & no & heavy \\
             & [1.1] & [1.1] & [4.0] & [1.1] \\
NH$_3$/N$_2$H$^+$ central enhancement  & yes & yes & ... & yes \\
             & [1.1] & [1.1] & ... & [either] \\
\hline
\multicolumn{5}{c}{Kinematics}\\
\hline
Non-thermal Line width [km s$^{-1}$] &
0.17\tablenotemark{b} &
0.17\tablenotemark{b} &
0.27\tablenotemark{c} &
0.2\tablenotemark{b} \\
Infall Velocity [km s$^{-1}$] & 0.03-0.044\tablenotemark{d} &
   &
   & 0.021-0.1\tablenotemark{d,e} \\
                & [1.1] & [1.1] & [4.0] & [1.1] \\
\enddata

\tablenotetext{a}{A preferred value of $\alpha$}
\tablenotetext{b}{\citet{taf2002}}
\tablenotetext{c}{\citet{ts2004}}
\tablenotetext{d}{\citet{lee2001}}
\tablenotetext{e}{\citet{oha1999}}

\end{deluxetable}

\subsection{Formation of Cores and Role of Turbulence}
The line widths of molecular cloud tracers are typically significantly larger
than the thermal line width, which suggests that the clouds are dominated
by non-thermal motions often identified as turbulence \citep[e.g.][]{mg1988,
fg1992, gb1998, bb2000}.
The existence and dissipation of turbulence are considered to be one of the
keys controlling star formation \citep[e.g.][]{bkv2003,
mk2004}. In our present paper, we do not consider
turbulence. Since observed molecular line widths in many prestellar cores are
dominated by thermal motions \citep{taf2002, tuk2004},
this simplification is justified at least for those objects.

However, the different values of $\alpha$ in our models can tell us something
about the time scale for the ambient material to be
accumulated onto the core and for turbulence to dissipate.
Although the origin of turbulence is not well understood, turbulence
will decay within a sound-crossing time scale, without excitation
sources \citep[e.g.][]{sog1998}.
If the accumulation of interstellar gas is efficient and a core becomes too
massive to be supported by thermal pressure before turbulence dissipates,
it collapses dynamically after the dissipation (i.e. resulting in large
$\alpha$). On the other hand, if the accumulation is rather slow and
turbulence dissipates before the core mass reaches the critical
value, the core becomes a stable Bonnor-Ebert sphere. Additional accumulation
will achieve a critical Bonnor-Ebert core with $\alpha\sim 1$.

As discussed in the previous subsection (see also \S 3.1 and \S 3.2),
both the line width and molecular distributions
obtained in our model with $\alpha=1.1$ are consistent with the observations
in L1517B and L1498.
On the other hand, molecular distributions in L1521E are better reproduced by
the $\alpha=4.0$ model. Since observed line widths are relatively large,
as predicted in the model, L1521E could  indeed be a core with large $\alpha$.
However, a clear infall signature is not observed in this object.
Based on  the above core  formation processes, one  possibility is
that  L1521E  has  accumulated  enough  mass to  be  recognized  as  a
prestellar  core before the dissipation of the turbulence. In this case,
the turbulence accounts for the relatively large line width, and the
relatively fast accumulation (before the turbulence decays) accounts for the
{\it young} ($\lesssim 10^5$ yr) molecular abundances with little or no
depletion.
Detailed observation of velocity fields in chemically young cores
such as L1521E and studies on the origin of turbulence in molecular clouds
are highly desirable in order to understand the formation of cores.

\section{Summary}
We have utilized numerical calculations to investigate the gravitational
contraction of spherical cloud cores and the molecular distribution
synthesized within these collapsing cores.

Our major findings are that:
\begin{itemize}
\item{The observed infall
velocities in prestellar cores, typically $\sim 0.05-0.1$ km s$^{-1}$
when the central density is $10^5$ cm$^{-3}$, are reproduced by the
contraction of a core that is initially close to the critical Bonnor-Ebert
sphere ($\alpha = 1.1$) with a central density of $2 \times 10^4$
cm$^{-3}$. Note that $\alpha$ is the ratio of the gravitational force to
the pressure force, and $\alpha = 1.0$ corresponds to the equilibrium sphere.}

\item{When the central density of the core model with $\alpha=1.1$ reaches
$n_{\rm H}=3\times 10^{5}$ cm$^{-3}$, carbon-bearing species are significantly
depleted in the central region, which is consistent with observations of
L1517B and L1498.}

\item{In the model with $\alpha=4.0$, the contraction is faster, and hence the
depletion of molecules is less significant than in the model with
$\alpha=1.1$. When the central density is $3 \times 10^5$ cm$^{-3}$, the
molecular distribution is consistent with that in L1521E. The molecular
distribution can thus be a probe of the contraction or accumulation time scale
of cores.}

\item{The central enhancement of the NH$_3$ to N$_2$H$^+$ ratio,
which is observed in prestellar cores, is reproduced at a central
core density of $3 \times 10^{5}$ cm$^{-3}$ by adopting a recently
measured branching ratio for the N$_2$H$^+$ dissociative recombination.}

\item{The column densities of carbon-bearing species are higher in the model
with larger $\alpha$, and they tend to vary little with time when the central
density of the core is $3\times 10^5-3\times 10^7$ cm$^{-3}$. On the other
hand, the column densities of N-bearing species and deuterated species tend
to increase with contraction.}

\item{Significant deuterium fractionation is obtained, especially
when $\alpha=1.1$. Particularly striking are the D/H
ratios of deuterated isotopomers of H$_3^+$, which lie similar 
to or higher than unity at the
core center when the central density is $\gtrsim 3\times 10^6$ cm$^{-3}$.
The deuterated isotopomers of H$_3^+$ can be useful in probing the
central regions of evolved cores in which species containing heavy
elements are depleted from the gas.}

\item{In the realm of neutral multi-deuterated species, our model is capable
of reproducing the observed abundance ratio of ND$_3$/NH$_3$ found
in protostellar objects, but underestimates the deuterium fractionation
of formaldehyde and methanol observed in prestellar cores and the
protostellar object IRAS 16293.}

\item{The composition of ice mantles depends sensitively
on the temperature. While CO ice is dominant in the model with $T=10$ K,
CH$_3$OH and CO$_2$ are the most abundant C-bearing species in  ice mantles
for the models with $T=12$ K and 15 K, respectively. In reality, temperature
varies with space and time within prestellar cores. Consideration of such
temperature variation is important for predicting the ice composition.
Evaluation of relevant grain-surface processes in laboratory experiments
and comparison between the modified-rate  and stochastic methods using a
large grain-surface-reaction network are highly desirable.}

\end{itemize}

\acknowledgments

We are grateful to the members of the CANS group for providing an original
code to calculate the one-dimensional collapse model. We especially thank
Drs. Fukuda and Yokoyama for their help in developing and tuning the code.
We thank Dr. Stantcheva for helpful discussions on grain-surface processes.
We are grateful to the referee for helpful comments.
Y. A. is supported by a Grant-in-Aid for Scientific Research (14740130,
16036205) and ``The 21st Century COE Program of Origin and Evolution of
Planetary Systems" of the Ministry of Education, Culture, Sports, Science
and Technology of Japan (MEXT).
The Astrochemistry program at The Ohio State
University is supported by the National Science Foundation.  P. C.
acknowledges support from the MIUR
project ``Dust and Molecules in Astrophysical Environments.''  Numerical
calculations were carried out on the VPP5000 at the Astronomical Data
Analysis Center of the National Astronomical Observatory of Japan.

\clearpage


\end{document}